\documentclass[aps, twocolumn,superscriptaddress]{revtex4-2} \usepackage{natbib} \usepackage[colorlinks=true]{hyperref}

\usepackage{bm}
\usepackage{physics}
\usepackage[normalem]{ulem}
\usepackage{mathtools}
\usepackage{amsfonts}
\usepackage{amssymb}
\usepackage{dsfont}
\newcommand{\rAngle}{\rangle \hspace{-2pt} \rangle }
\newcommand{\lAngle}{\langle \hspace{-2pt} \langle }
\hypersetup{
    unicode=false,          
    pdftoolbar=true,        
    pdfmenubar=true,        
    pdffitwindow=false,     
    pdfstartview={FitH},    
    pdfsubject={},          
    pdfcreator={},          
    pdfproducer={},         
    pdfkeywords={} {} {},   
    pdfnewwindow=true,      
    colorlinks=true,        
    linkcolor=magenta,      
    citecolor=blue,         
    filecolor=magenta,      
    urlcolor=blue           
} 
\allowdisplaybreaks

\renewcommand{\bar}{\overline}

\renewcommand{\hat}{\widehat}
\renewcommand{\leq}{\leqslant}
\renewcommand{\geq}{\geqslant}

\renewcommand{\Im}{\operatorname{Im}}
\newcommand{\pdagger}{{\phantom{\dagger}}}
\newcommand{\equref}[1]{Eq.~(\ref{#1})}
\newcommand{\id}{\mathds{1}}

\begin{document}
\title{An exactly solvable dissipative spin liquid}
\author{Leyna Shackleton}
\affiliation{Department of Physics, Harvard University, Cambridge MA 02138, USA}
\author{Mathias S.~Scheurer}
\affiliation{Institute for Theoretical Physics III, University of Stuttgart, 70550 Stuttgart, Germany}
\affiliation{Institute for Theoretical Physics, University of Innsbruck, Innsbruck A-6020, Austria}
\date{January 22, 2024\vspace{0.4in}}

\begin{abstract}
Exactly solvable Hamiltonians with spin liquid ground states have proven to be extremely useful, not only because they unambiguously demonstrate that these phases can arise in systems of interacting spins but also as a pedagogical illustration of the concept and as a controlled starting point for further theoretical analysis. However, adding dissipative couplings to the environment---an important aspect for the realization of these phases---generically spoils the exact solvability. We here present and study a Lindbladian, describing a square-lattice spin-liquid with dissipative coupling to the environment, that admits an exact solution in terms of Majorana fermions coupled to static $\mathbb{Z}_2$ gauge fields. This solution allows us to characterize the steady-state solutions as well as ``quasiparticle'' excitations within the Lindbladian spectrum. We uncover distinct types of quasiparticle excitations of the Lindbladian associated with parametrically different timescales governing the equilibration time of the expectation values of different classes of observables. Most notably, for small but non-zero dissipation, we find a separation into three different timescales associated with a three-step heating profile. 
On a more general level, our exactly solvable Lindbladian is expected to provide a starting point for a better understanding of the behavior of fractionalized systems under dissipative time evolution.
\end{abstract}

\maketitle
\section{Introduction}

Quantum spin liquids (QSLs) are exotic phases of matter characterized by emergent anyon excitations with non-trivial braiding statistics, in conjunction with the absence of any conventional long-range order~\cite{kitaev2003, savary2016, broholm2020}. Further interest in these states have grown due to their potential applications for use in fault-tolerant quantum computation~\cite{dennis2002, terhal2015} through their non-local encoding of quantum information.

The interplay between QSLs and open quantum systems has been an active area of research for many years, with a primary focus on the robustness of their information storage and on approaches to detect their presence when perturbations generic to experimental realization are introduced, such as a non-zero temperature, decoherence, and more~\cite{nussinov2008, alicki2009, brown2016, PhysRevB.99.075141, self2019,PhysRevX.12.041004,PhysRevResearch.3.L032024,2022arXiv221109784Z}. Rather than taking this approach of considering generic forms of decoherence, we instead consider engineering a particular form of environmental coupling to a QSL in order to realize unique non-equilibrium physics. This general approach of leveraging dissipation has been shown to be efficient at preparing quantum states~\cite{kraus2008,OpenQuantumPrepare,RydbergHanspeter,TrappedIons,PhysRevA.92.012128} including topologically-protected edge modes~\cite{diehl2011}. Recent applications of this idea to spin liquids~\cite{yang2021, hwang2023} have yielded new insights into the behavior of emergent anyon excitations in the presence of dissipation.

We study a quantum spin-$3 / 2$ model on a two-dimensional square lattice, which is a particular limit of the QSL studied in~\cite{yao2009}, and subject it to a certain choice of Markovian open dynamics generated by the Lindblad equation. We show that in a particular limit, the Lindbladian becomes exactly solvable through a parton construction. As such, exact statements about its steady-state solutions as well as transient behavior can be made. Exactly solvable Lindbladians have been studied previously using techniques such as third quantization~\cite{prosen2008,PhysRevResearch.3.033022,2023arXiv230401836S}, Bethe ans\"atze~\cite{medvedyeva2016, deleeuw2021}, operator-space fragmentation~\cite{essler2020}, and through parton constructions~\cite{shibata2019} similar to our own. From a practical perspective, this exact solvability is especially useful as the wealth of analytic tools developed to approximately study the low-energy behavior of Hermitian Hamiltonians do not immediately carry over to these non-Hermitian Lindbladians, although several methods for approximately studying the spectrum of Lindbladians have been developed~\cite{reiter2012, kessler2012}.

A particular property of our exact solution that we emphasize is the existence of distinct quasiparticle excitations of the Lindbladian when viewed as an effective non-Hermitian Hamiltonian acting on an enlarged Hilbert space. We advocate for this as a powerful tool for understanding the non-equilibrium behavior of a generic state or density matrix as it equilibrates to its steady-state solution. We show that the imaginary energy gap associated with a particular type of quasiparticle excitation in this enlarged Hilbert space can be associated with the equilibration timescale of the expectation value of a certain class of observables. These classes of observables turn out to have a close relation to excitations of the corresponding unitary spin liquid. 
An expert reader might immediately want to inspect Sec.~\ref{sec:summary} for a summary of the spectrum. Importantly, the different time scales of these classes of operators have different parametric dependence on the strength $\gamma$ of the coupling to the environment, which can be found simply by diagonalizing a quadratic Hamiltonian numerically, or in some cases is derived exactly analytically. For instance, in the limit of small $\gamma$, a certain set of operators, that are not conserved by the unitary dynamics, decay rapidly on a scale set by the exchange coupling rather than $\gamma$ itself. Fractionalized string-like operators that can be interpreted as pairs of emergent Majorana fermion excitations in the unitary system, however, survive up to a time-scale $\propto 1/\gamma$. After that, also the Majorana fermions heat up and only gauge-invariant fluxes of the emergent gauge fields or Wilson-loop operators remain in their original configuration. In this sense, our model realizes a three-step and exactly solvable analogue of the ``fractionalized pre-thermalization'' discussed recently \cite{jin2022} for stroboscopic time-evolution in the Kitaev model.

The remainder of the paper is organized as follows. A mathematical definition of all the involved operators and of the dissipative model we study can be found in Sec.~\ref{sec:model}. We derive an interpret the spectrum of the Lindbladian in Sec.~\ref{sec:spectrum}. A discussion of perturbations away from the exactly solvable point and a conclusion are provided in Sec.~\ref{sec:perturbations} and Sec.~\ref{sec:summary}, respectively.

\section{Model}
\label{sec:model}
The time evolution of a density matrix $\rho$ can be described in its most general form by a completely-positive and trace preserving map $\Phi(\rho) \rightarrow \rho'$.
The Lindblad equation~\cite{gorini1976, lindblad1976} is the most generic continuous Markovian map satisfying these properties,
\begin{equation}
  \begin{aligned}
    \dv{\rho}{t} &=  \mathcal{L}\left[ \rho \right] = - i \comm{H}{\rho} + \sum_j \left( L_j \rho L_j^\dagger - \frac{1}{2}\acomm{L_j^\dagger L_j^\pdagger}{\rho} \right) \,,
  \end{aligned}
\end{equation}
where the quantum jump operators $L_j$ parameterize the nature of the environmental coupling. One may express the superoperator $\mathcal{L}$ as an operator in a ``doubled'' Hilbert space, namely the Hilbert space of all operators. For a choice of basis in the original Hilbert space, $\ket{\psi_i}$, $i = 1\ldots \mathcal{D}$, we can represent any operator $\mathcal{O} = \sum_i \mathcal{O}_{ij} \ket{\psi_i}\bra{\psi_j}$ as a state $\Vert \mathcal{O} \rAngle \equiv \sum_{ij} \mathcal{O}_{ij} \ket{\psi_i} \otimes \ket{\psi_j}$ in this doubled Hilbert space, with inner product $\lAngle \mathcal{O}_1 \Vert \mathcal{O}_2 \rAngle = \frac{1}{\mathcal{D}} \tr \left( \mathcal{O}_1^\dagger \mathcal{O}_2 \right)$. Within this doubled Hilbert space, the action of the Lindbladian superoperator is
\begin{equation}
  \begin{aligned}
    i \mathcal{L}   &= H_{\text{eff}} \otimes \id - \id \otimes H_{\text{eff}}^\dagger + \sum_j i \gamma L_j \otimes L_j^\dagger\,,
    \\
    H_{\text{eff}} &\equiv H - \frac{i \gamma}{2} \sum_j L_j^\dagger L_j^\pdagger \,.\label{GeneralBilayerModel}
  \end{aligned}
\end{equation}
We will take $L_j$ to be unitary, such that $H_{\text{eff}} = H$ up to an overall imaginary constant. 

This doubled Hilbert space construction is a powerful tool for characterizing the behavior of mixed states; notably, it has seen recent use in diagnosing the stability of quantum information stored in mixed states~\cite{bao2023, lee2023}. For a quantum spin model in two dimensions, it is instructive to think of this doubled Hilbert space as corresponding to a bilayer system, where the first (second) layer corresponds to the bra (ket). In this scenario, the Lindbladian consists of two copies of the Hamiltonian $\pm H$ acting on each of the two layers, with anti-Hermitian couplings $i \gamma \sum_j L_j \otimes L_j^\dagger$ between the two layers. To better connect with intuition from unitary time evolution, we will focus on the eigenvalues of the matrix $i\mathcal{L}$ rather than $\mathcal{L}$ and refer to $i\mathcal{L}$ as ``the Lindbladian''; in this convention, the imaginary components of eigenvalues correspond to dissipation, and the non-existence of exponentially growing solutions requires the imaginary part to always be negative.

\subsection{Unitary time evolution}
The Hermitian dynamics that we consider is a particular limit of an exactly solvable quantum spin-$3 / 2$ model on a square lattice first studied in~\cite{yao2009}. We define this model here and review some properties of its solution, as our results are most clearly stated within this framework. Due to the four spin polarizations per site, we may express the spin-$3 / 2$ degrees of freedom in terms of anticommuting Gamma matrices $\Gamma^a$, $a = 1\ldots 5$, which obey $\acomm{\Gamma^a}{\Gamma^b} = 2 \delta^{ab}$. In terms of the physical spin operators,
\begin{equation}
  \begin{aligned}
    \Gamma^1 &= \frac{1}{\sqrt{3}} \acomm{S^y}{S^z}\,, \quad \Gamma^2 = \frac{1}{\sqrt{3} }\acomm{S^z}{S^x}\,,  \\
    \Gamma^3 &= \frac{1}{\sqrt{3}} \acomm{S^x}{S^y} \,, \quad \Gamma^4 = \frac{1}{\sqrt{3}} \left[ (S^x)^2 - (S^y)^2 \right] \,, \\
    \Gamma^5 &= (S^z)^2 - \frac{5}{4}\,. \\
  \end{aligned}
\end{equation}
We emphasize that the key property needed in our construction is the presence of five anti-commuting Gamma matrices. This can alternatively be accomplished by a pair of spin-$1/2$ operators (or qubits) on each site. In this approach, there are multiple ways of constructing anti-commuting Gamma matrices. One possible representation is
\begin{equation}
  \begin{aligned}
    \Gamma^1 &= S^x \otimes S^x \,, \quad \Gamma^2 = S^x \otimes S^y \,,  \\
    \Gamma^3 &= S^x \otimes S^z \,, \quad \Gamma^4 = S^y \otimes \id \,, \\
    \Gamma^5 &=S^z \otimes \id \,. \\
  \end{aligned}
\end{equation}
The choice of representation will influence the physical interpretation of the dissipation, as will be discussed later. Additional choices are discussed in Appendix~\ref{app:gamma}.

The Hamiltonian is defined on a square lattice as
\begin{equation}
  \begin{aligned}
  H &= \sum_j \left[ J_x \Gamma^1_j \Gamma^2_{j + \hat{x}} + J_y \Gamma^3_j \Gamma^4_{j + \hat{y}} \right]  \\
  &+ \sum_j \left[ J_x' \Gamma^{15}_j \Gamma^{25}_{j + \hat{x}} + J_y' \Gamma^{35}_j \Gamma^{45}_{j + \hat{y}}\right] - J_5 \sum_j \Gamma^5_j
  \label{eq:squareHamiltonian}
  \end{aligned}
\end{equation}
where $\Gamma^{ab}_j \equiv \comm{\Gamma^a_j}{\Gamma^b_j} / 2 i$. For simplicity, we will assume that the lattice has an even number of sites in both the $\hat{x}$ and $\hat{y}$ directions. The exact solvability of this model is a consequence of an extensive number of conserved fluxes, 
\begin{equation}
    W_j = \Gamma^{13}_j \Gamma^{23}_{j + \hat{x}} \Gamma^{14}_{j + \hat{y}} \Gamma^{24}_{j + \hat{x} + \hat{y}}, \label{WOperators}
\end{equation}
and can be understood most conveniently by performing a Majorana decomposition of the $\Gamma$ matrices; specifically, one employs the representation
\begin{equation}
  \begin{aligned}
    \Gamma_{j}^\mu &= i c_{j}^\mu d_{j} \,, \quad \Gamma_{j}^{\mu 5} = i c_{j}^\mu d_{j}' \,, \quad \mu = 1\,, 2\,, 3\,, 4\,, 
    \\
    \Gamma_{j}^5 &= i d_{j} d_{j}'\,, \label{DefinitionOfMajoranas}
  \end{aligned}
\end{equation}
with the constraint $- i c_{j}^1 c_{j}^2 c_{j}^3 c_{j}^4 d_{j} d_{j}' = \Gamma_{j }^1 \Gamma_{j}^2 \Gamma_{j}^3 \Gamma_{j}^4 \Gamma_{j}^5 = -1$. In this representation, the Hamiltonian can be rewritten in terms of static $\mathbb{Z}_2$ gauge fields $\hat{w}_{j,\alpha}$ living on the bonds of the lattice, which come from conserved bilinears of the $c_j^\mu$ operators, coupled to two species of Majorana fermions, $d_j$ and $d_j'$. 

We will not give a detailed review of the various properties of this solution~\cite{yao2009}, as it will not be important for our analysis. However, we will emphasize the relation between these emergent degrees of freedom and physical observables, as the results of our dissipative model concisely fit into this picture. The $\mathbb{Z}_2$ gauge fluxes - products of closed loops of $\hat{w}_{j,\alpha}$ operators - correspond to the conserved fluxes $W_j$. Pairs of Majorana fermions coupled by a string of $\mathbb{Z}_2$ gauge fields are given by strings of $\Gamma$ matrices. For a pair of $d$ excitations, the operator can be generated by a string of bond operators:
\begin{equation}
  \begin{aligned}
    V_{j, \alpha} &= \begin{cases}
    \Gamma^{1}_j \Gamma^{2}_{j + \hat{x}} & \alpha = x\,, \\
    \Gamma^{3}_j \Gamma^{4}_{j + \hat{y}} & \alpha = y\,. \\
  \end{cases}
  \label{eq:bondOperators1}
  \end{aligned}
\end{equation}
A similar construction follows for a pair of $d'$ fermions,
\begin{equation}
  \begin{aligned}
    V_{j ,\alpha}' &= \begin{cases}
    \Gamma^{15}_j \Gamma^{25}_{j + \hat{x}} & \alpha = x\,, \\
    \Gamma^{35}_j \Gamma^{45}_{j + \hat{y}} & \alpha = y\,, \\
  \end{cases}
  \label{eq:bondOperators}
  \end{aligned}
\end{equation}
as well as the combination of a $d$ and $d'$ fermion, a special case of which is $\Gamma^5_j = i d_j d_j'$.
Note that a closed loop of either the $V_{j,\alpha}$ or $V_{j,\alpha}'$ operators is equivalent to a product of the conserved fluxes contained inside the loop.

In order to retain the exact solvability upon the inclusion of dissipation, we take $J_x' = J_y' = J_5 = 0$, which causes the bond operators $V_{j ,\alpha}'$ to become conserved quantities. In the Majorana fermion language, this limit quenches the dispersion of the $d_j'$ fermions and the ground state becomes highly degenerate as pairs of $d_j'$ may be added in at no energy cost.

\subsection{Jump operators}
We now introduce jump operators $L_j = \Gamma^5_j$. Note that our Lindbladian jump operators commute with the conserved flux, $\comm{L_j}{W_k} = 0$. This property implies that the flux operators $W_j$ constitute \textit{strong symmetries} of the system, as defined in~\cite{buca2012}, and means that an initial state with a definite flux configuration will remain in such a configuration. If we express our Hermitian model as free Majorana fermions coupled to a static $\mathbb{Z}_2$ gauge field, the interpretation of this phenomenon is that the gauge fields will remain static under the Lindbladian time evolution while generically we expect the Majorana fermions to evolve to resemble a finite-temperature Gibbs state. One may think of this behavior as ``fractionalized thermalization.'' For a generic set of quantum jump operators that commute with $W_j$, we expect the steady-state solutions of the Lindbladian can be represented as the tensor product of a thermal Gibbs state of Majorana fermions with a pure state of $\mathbb{Z}_2$ gauge fields. We note related work studying the separation of thermalization timescales in fractionalized excitations on the Kitaev honeycomb model~\cite{jin2022} under stroboscopic time evolution, as well as more directly analogous work studying the Kitaev honeycomb model coupled to jump operators that commute with the conserved fluxes~\cite{hwang2023}. Apart from fluxes being exactly conserved under dissipative dynamics, we also uncover below an additional, less apparent regime of fractionalized thermalization in our exactly solvable model, which occurs in the limit of small dissipation.

\begin{figure}[tb]
  \centering
  \includegraphics[width=0.45\textwidth]{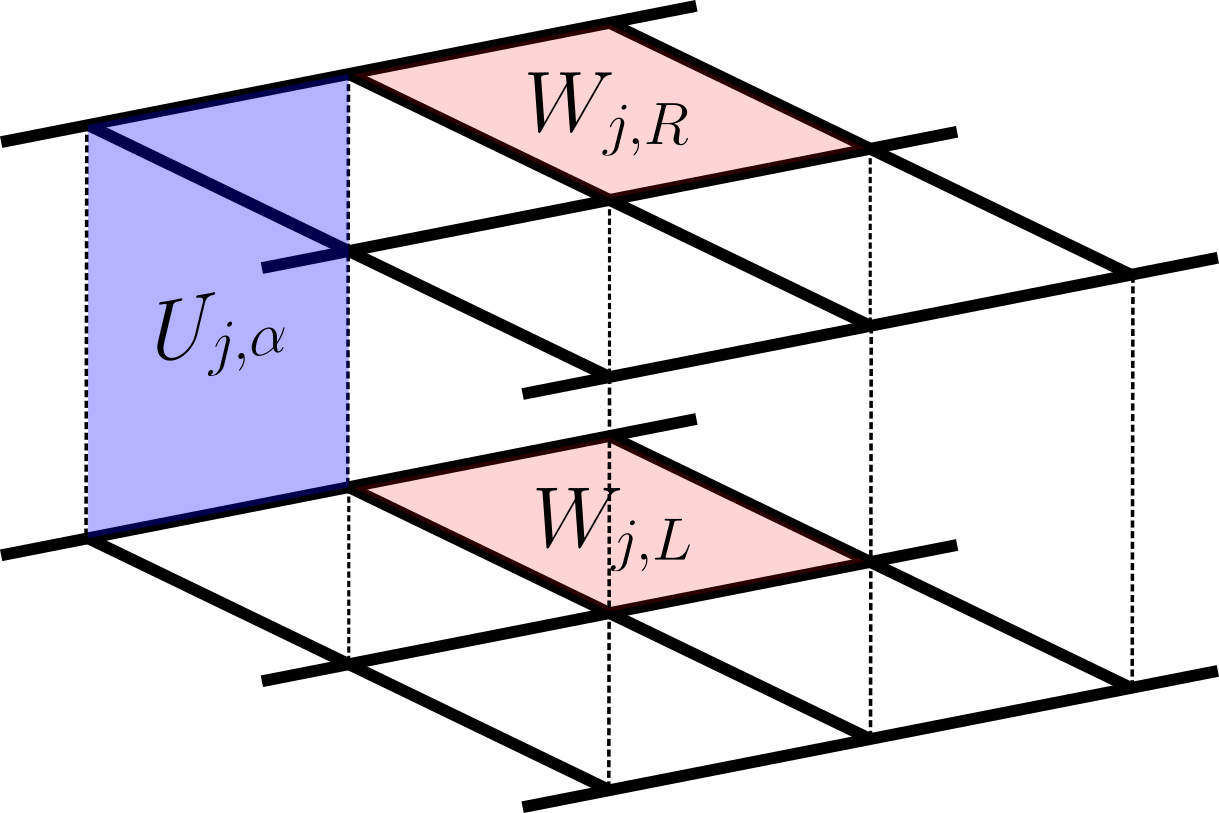}
  \caption{In the doubled Hilbert space representation, the Lindbladian super operator possesses two types of conserved fluxes. The first are intralayer fluxes $W_{j,R}$, $W_{j,L}$, which correspond to physical conserved plaquette operators. The second, $U_{j,\alpha}$, have a purely superoperator interpretation, as explained in the main text.}
  \label{fig:bilayerFlux}
\end{figure}

The above discussion follows for any jump operator that commutes with the conserved fluxes, and remains true even away from the limit $J'_x = J'_y = J_5 = 0$. However, our particular model admits additional conserved quantities which render the full dissipative dynamics exactly solvable.
To see this, we use the doubled Hilbert space formalism, see \equref{GeneralBilayerModel}, to express the Lindbladian superoperator as an operator acting on a bilayer spin-$3/2$ system, with Gamma matrices $\Gamma^a_R\,, \Gamma^a_L$ for the two layers - the $R\,, L$ subscript indicates that they correspond to the right and left action of the gamma matrices on the physical operator. The Lindbladian can be written as
\begin{equation}
  \begin{aligned}
    i \mathcal{L} = H[ \Gamma_R ] - H[ \Gamma_L ] + i \gamma \sum_j \Gamma_{j, R}^5 \Gamma_{j,L}^5  - i \gamma N,
    \label{eq:lindbladianGamma}
  \end{aligned}
\end{equation}
where $N$ is the number of sites.
This bilayer representation makes it clear that, in addition to the intralayer fluxes $W_{j, R}\,, W_{j ,L}$ which are defined in analogy to \equref{WOperators} and commute with the Lindbladian separately, we have a new set of conserved interlayer fluxes $U_{j,\alpha} \equiv V_{j, \alpha, R}' V_{j,\alpha, L}'$ defined on the plaquettes connecting the two layers, shown in Fig.~\ref{fig:bilayerFlux}. These conserved quantities are ``weak'' symmetries~\cite{buca2012}. In contrast to the strong symmetries generated by the flux operators $W_j$, the operators $V_{j ,\alpha}'$ do not commute with the jump operators $L_j$ individually, and it is exclusively the conserved \textit{superoperator} consisting of the simultaneous right and left action of $V_{j,\alpha}'$ that commutes with the Lindbladian. 

We comment here on the physical interpretation of the jump operators $\Gamma^5_j$ in terms of the microscopic degrees of freedom. If our Gamma matrices are built out of pairs of spin-$1/2$ operators, $\Gamma^5 = S^z \otimes \id$ and our dissipation should be thought of as an \textit{asymmetric} dephasing acting on only one of the two spin-$1/2$ degrees of freedom. For spin-$3/2$ operators, $\Gamma^5 = (S^z)^2 - \frac{5}{4}$, which acts as a dephasing term between the $S^z = \pm \frac{1}{2}$ and the $S^z = \pm \frac{3}{4}$ states. 
\subsection{Parton construction}
To elucidate the exact solvability of this model, we represent the Gamma matrices in terms of six Majorana fermions,
\begin{equation}
  \begin{aligned}
    \Gamma_{j, R}^\mu &= i c_{j, R}^\mu d_{j, R} \,, \quad \Gamma_{j, R}^{\mu 5} = i c_{j, R}^\mu d_{j, R}' \,, \quad \mu = 1\,, 2\,, 3\,, 4\,, 
    \\
    \Gamma_{j, R}^5 &= i d_{j, R} d_{j, R}'\,,
  \end{aligned}
\end{equation}
with an analogous representation for $\Gamma^\mu_L$ in terms of $c_{j, L}^\mu \,, d_{j, L} \,, d_{j, L} ' $. This enlarges our Hilbert space, which necessitates the constraint $- i c_{j, R}^1 c_{j, R}^2 c_{j,R}^3 c_{j, R}^4 d_{j, R} d_{j, R}' = \Gamma_{j, R}^1 \Gamma_{j, R}^2 \Gamma_{j, R}^3 \Gamma_{j, R}^4 \Gamma_{j, R}^5 = -1$ on all physical states, and likewise for the $\Gamma_L$ operators.

In this representation, the Hamiltonian $H[\Gamma_R]$ becomes
\begin{equation}
  \begin{aligned}
    H[\Gamma_R] = \sum_j J_x \hat{w}_{j, x, R} i d_{j, R} d_{j + \hat{x}, R} + J_y \hat{w}_{j, y, R} i d_{j, R} d_{j + \hat{y}, R}
  \end{aligned}
\end{equation}
where $\hat{w}_{j, x, R} \equiv -i c_{j, R}^1 c^2_{j + \hat{x}, R}$ and $\hat{w}_{j, y, R} \equiv - i c_{j, R}^3 c^4_{j + \hat{y}, R}$ are conserved quantities with eigenvalue $\pm 1$. An analogous rewriting follows for the Hamiltonian on the second layer. Observe that the Majorana fermions $d_{j, R}'\,, d_{j, L}'$ drop out of the intralayer Hamiltonian entirely. As a result, the interlayer coupling also becomes quadratic in the Majorana fermions,
\begin{equation}
  \begin{aligned}
    &i \gamma \sum_j \Gamma^5_{j, R} \Gamma^5_{j, L} = - i \sum_j d_{j, R} d_{j, R}' d_{j, L} d_{j, L}' 
    \\
    &= - \gamma \sum_j \hat{v}_j d_{j, R} d_{j, L}\,,
  \end{aligned}
\end{equation}
where $\hat{v}_j \equiv -i d_{j, R}' d_{j, L}'$ is a conserved quantity with eigenvalue $\pm 1$. With this rewriting, our model becomes one of free fermions $d_{j, R}\,, d_{j, L}$ hopping on a bilayer square lattice in the presence of a background $\mathbb{Z}_2$ gauge field $\hat{w}_{j, \alpha, R}\,, \hat{w}_{j,\alpha, L}\,, \hat{v}_j$ living on the links. Written out explicitly, 
\begin{widetext}
\begin{equation}
  \begin{aligned}
  i \mathcal{L} &= \sum_{\ell = L, R} \sum_j s_\ell \left[ J_x  \hat{w}_{j, x, \ell} j d_{j, \ell} d_{j + \hat{x}, \ell} + J_y  \hat{w}_{j, y, \ell} i d_{j, \ell} d_{j + \hat{y}, \ell}  \right] - \gamma \sum_j \hat{v_j} d_{j, R} d_{j, L}  - i \gamma N
  \label{eq:freeFermionHamiltonian}
  \end{aligned}
\end{equation}
\end{widetext}
where $s_L = 1$, $s_R = -1$. This Lindbladian possesses a local $\mathbb{Z}_2$ gauge symmetry, given by the transformation $d_{j, \ell} \rightarrow \Lambda_{j, \ell} d_{j, \ell}$, $\hat{w}_{j, \alpha, \ell} \rightarrow \Lambda_{j, \ell} \hat{w}_{j, \alpha, \ell} \Lambda_{j + \hat{\alpha}, \ell}$, $\hat{v}_j \rightarrow \Lambda_{j, L} \hat{v}_j \Lambda_{j, R}$, where $\Lambda_{j, \ell} = \pm 1$. The gauge-invariant fluxes around a single intralayer plaquette gives the conserved quantities $-W_{j, R}\,, -W_{j, L}$, and the fluxes around an interlayer plaquette gives the conserved superoperator $-U_{j,\alpha}$. Note the relative minus signs between the two quantities - as will be relevant later, working in a sector with $U_{j,\alpha} = 1$, which is the sector where steady-state solutions will belong to, requires us to pick a gauge configuration such as $\hat{v}_j = (-1)^{j}$. 

In order to obtain physical states, we must project back to our physical (doubled) Hilbert space. This is obtained by the projection operator $P = \prod_{j, \ell} \frac{1 + D_{j, \ell}}{2}$, where $D_{j, \ell} = - i c^1_{j, \ell} c^2_{j, \ell} c^3_{j, \ell} c^4_{j, \ell} d_{j, \ell} d'_{j, \ell}$. A careful analysis of this for a single-layer Hamiltonian was performed in~\cite{yao2009} and our analysis proceeds along similar lines. We can write $P = P'(1+D)$, where $D \equiv \prod_{j, \ell} D_{j, \ell}$ and $P'$ is a linear combination of all inequivalent gauge transformations. Since $D^2 = 1$, $\comm{D}{\mathcal{L}} = 0$, this means that we must restrict ourselves to eigenstates with $D = 1$. We write
\begin{equation}
  \begin{aligned}
D = \prod_{j, \alpha, \ell} \hat{w}_{j, \alpha, \ell} \prod_j \hat{v}_j \prod_j i d_{j, L} d_{j, R}\,.
\label{eq:gaugeInvariance}
  \end{aligned}
\end{equation}
In order to more readily leverage the gauge constraint, we re-express the Majorana fermions $d_{j, R}\,, d_{j, L}$ in terms of complex fermions. A representation that will prove to be useful for future analysis is
\begin{equation}
  \begin{aligned}
  f_j = i^j \left( d_{j, L} + i (-1)^{j} d_{j, R} \right) / 2 \,.  
  \label{eq:complexRep}
  \end{aligned}
\end{equation}
With this, $2 f_j^\dagger f_j - 1 = (-1)^{j} i d_{j, L} d_{j, R}$ and $(-1)^{N_f} \equiv (-1)^{\sum_j f_j^\dagger f_j} = \prod_j i d_{j, L} d_{j, R}$. Therefore, gauge invariance restricts the total fermion parity, $(-1)^{N_f}$, to equal the total ``gauge parity,'' $\prod_{j, \alpha, \mu} \hat{w}_{j, \alpha, \mu} \prod_j \hat{v}_j$.

\section{Spectrum of the Lindbladian}%
\label{sec:spectrum}
In the previous section, we have shown that our Lindbladian reduces down to one of free fermions coupled to a static $\mathbb{Z}_2$ gauge field. As such, the full spectrum and eigenvectors can in principle be calculated - analytically for translationally-invariant gauge field configurations, and by diagonalizing a non-Hermitian single-particle Hamiltonian for more general gauge configurations. However, the interpretation of these properties must be done in terms of density matrices of our physical Hilbert space, rather than a more conventional analysis of Hermitian systems. We outline our general approach to understanding these properties below.

\subsection{General remarks}
\label{subsec:generalRemarks}
The most important eigenstates of the Lindbladian are those with eigenvalue zero, which correspond to steady-state solutions. Since the eigenvalues $\lambda_i$ of the Lindbladian obey $\Im[\lambda_i] \leq 0$, every initial density matrix will eventually evolve into some superposition of these steady-state solutions (for simplicity, we we ignore the possibility of solutions with purely real eigenvalue, i.e. density matrices that do not decay but whose phase oscillates in time, as these are not present in our spectrum). Our first task will be to find these steady-state solutions and understand their properties.

Ascertaining the properties of these steady-state solutions is a non-trivial task within the doubled Hilbert space formalism. Given a density matrix $\Vert \rho \rAngle$, the expectation value of a Hermitian operator $A$ is given by $\Tr\left[ A \rho \right] = \lAngle A \Vert \rho \rAngle$. As such, standard intuition for calculating observables of pure states in ordinary Hilbert spaces, $\bra{\psi} A \ket{\psi}$, is not applicable here. While it is possible to develop the machinery to perform such calculations, we instead proceed with a more intuitive symmetry-based analysis. The exact solvability of our model provides an extensive number of superoperators that commute with the Lindbladian, and hence $\Vert \rho \rAngle$ will be an eigenstate of them. By decomposing our Hilbert space into subspaces with definite eigenvalue under these superoperators, we can conclude that $\lAngle A \Vert \rho \rAngle$ must vanish unless the two have the same eigenvalue. In general, this symmetry analysis only gives us limited information about $\Vert \rho \rAngle$. However, the extensive number of conserved quantities makes this perspective especially powerful for our model, and we will find that only a small amount of additional analysis is required to fully characterize the steady-state solution. 

After characterizing the steady-state solutions, we will analyze the dissipative solutions - operators with eigenvalue $\lambda_i$ obeying $\Im \lambda_i < 0$. We will be interested in eigenvalues whose imaginary components have the smallest magnitude, which defines the \textit{Liouvillian gap}, and a corresponding timescale associated with the decay to the steady-state solution. As the spectrum of our Lindbladian has the interpretation of fermions coupled to a $\mathbb{Z}_2$ gauge field, we find it insightful to define distinct types of Liouvillian gaps depending on the nature of the excitation. For example, one may inquire into the Liouvillian gap with respect to fermionic excitations, or with respect to gauge excitations (visons). This is not an arbitrary labeling, the motivation for which ties back to our symmetry-based analysis of steady-state solutions. Excitations within a given sector will have different eigenvalues under the symmetries of our Lindbladian, and hence can be characterized by distinct classes of observables that have a non-zero overlap with these excitations. The corresponding Liouvillian gap for these excitations specify a timescale which governs the rate at which the expectation values for these classes of observables asymptote to their steady-state solutions. We note that a similar hierarchy of timescales was recently studied in random local Liouvillians~\cite{wang2020} and in fact observed in simulations on a quantum computer~\cite{sommer2021} - in this model, the separation of timescales was associated with differing spatial extents of operators. A symmetry-based analysis of the low-energy properties of the Lindbladian spectra has also been recently leveraged in Brownian circuits to construct an effective hydrodynamics description of the real-time dynamics~\cite{ogunnaike2023}  

To be more explicit with our perspective, consider a steady-state solution $\Vert \rho_{ss} \rAngle$ and a dissipative solution $\Vert a \rAngle$ which we interpret as a quasiparticle excitation of type $a$. A physical density matrix can be constructed by $\Vert \rho_d \rAngle \equiv \Vert \rho_{ss} \rAngle + c \Vert a \rAngle$, where $c$ is some constant chosen to ensure $\Tr \left[ \rho_{d}^2 \right] < 1$. This density matrix asymptotes to $\Vert \rho_{ss} \rAngle$ at late times but displays transient behavior dictated by $\Vert a \rAngle$ up to a timescale $t_a = -\Im[\lambda_a]^{-1}$. It is useful to characterize this operator $a$ in terms of observables $\left\{\mathcal{O}_a\right\}$ such that $\Tr \left[ \mathcal{O}_a a \right] \neq 0$, in which case one can say that the expectation value of observables $\mathcal{O}_a$ relax to their steady-state values with a timescale dictated by $t_{a}$ for the density matrix $\Vert \rho_d \rAngle$. Of course, a generic initial density matrix will be more complicated than $\Vert \rho_d \rAngle$; however, if $\Vert a \rAngle$ is the lowest-energy excitation that has a non-zero overlap with the observables $\mathcal{O}_a$, then $t_a$ provides an upper bound on the equilibration timescale for the expectation value of these observables.

The utility of this picture is contingent on the operators $\mathcal{O}_a$ having a sufficiently simple representation. As we will show, these different classes of observables are most conveniently stated in terms of fractionalized operators acting on the original Hilbert space, such as the bond operators in Eq.~\ref{eq:bondOperators1} and Eq.~\ref{eq:bondOperators}. In other words, we demonstrate a close connection between fractionalized excited states in the \textit{doubled} Hilbert space formalism and fractionalized operators in the \textit{physical} Hilbert space, with the imaginary energy of the former defining the equilibration timescale of expectation values of the latter.

We note an emerging body of work~\cite{mori2020, haga2021, lee2023, mori2023} which take a conceptually related stance to our own, which is that the Liouvillian gap does not solely determine the relaxation time of a dissipative system to its steady-state solution. These works noted that an anomalously small overlap between left and right eigenmodes of the Lindbladian - induced, for example, through a non-Hermitian skin effect - can enhance the relaxation time of the system. While we have verified numerically in our model that no such small overlap is present, we note the similarity with our work in that the structure of the \textit{eigenvectors}, rather than purely the energy gap, can qualitatively change the nature of the relaxation process. In our case, certain long-lived eigenmodes only have an effect on distinct classes of observables, which we identify by leveraging the extensive amount of symmetries present in our model.

\subsection{Steady-state solutions}
We now study the properties of the steady-state solutions. Recall that for isolated systems with similar Hamiltonians (free fermions coupled to static $\mathbb{Z}_2$ gauge fields), there is a theorem due to Lieb~\cite{lieb1994} for bipartite lattices that fixes the gauge flux sector in which the ground state resides in. In a similar spirit, we leverage general arguments given in~\cite{buca2012} that allow us to deduce gauge flux sectors which support steady-state solutions. 

A fact that we will use in this argument is that any dissipative eigenstate of the Lindbladian must have zero trace - if it had a non-zero trace, the dissipative nature implies that the trace would decay in time, contradicting the trace preservation of the Lindbladian time evolution. Hence, the search for steady-state solutions can be recast as a search for eigenstates with a non-zero trace. This comes with the caveat that we may miss steady-state solutions that happen to also have zero trace; however, we explicitly diagonalize the Lindbladian for a $4 \times 4$ lattice in each gauge sector and have found no such solutions.

We first constrain the interlayer fluxes $U_{j,\alpha}$, which constitute weak symmetries. Recall that the superoperator $U_{j,\alpha}$ acts on density matrices as $U_{j,\alpha}[\rho] = V_{j,\alpha}' \rho V_{j,\alpha}'$. An eigenstate of $U_{j,\alpha}$ with non-zero trace must have eigenvalue $1$, since unitarity and Hermiticity of $V_{j,\alpha}'$ implies $\Tr [\rho] = \Tr[V_{j,\alpha}' \rho V_{j,\alpha}']$. Hence, we will constrain ourselves to the $U_{j,\alpha} = 1$ sector.

We now turn to the ``strong'' symmetries $W_i$. A similar argument as the last paragraph implies that we must constrain ourselves to sectors where $W_j \rho W_j = \rho$. However, recall that in the doubled Hilbert space formulation, the right and left fluxes ($W_{j, R}$ and $W_{j, L}$) are conserved separately. Hence, our analysis only constrains the eigenvalues of $W_{j, R}$ and $W_{j, L}$ to be the same. This is actually not a new constraint - the product of fluxes around any closed surface must be $+1$, so the constraint that all $U_{j,\alpha} = +1$ automatically implies $W_{j, R} = W_{j, L}$. We will denote this choice of $W_{j, R}\,, W_{j, L}$ eigenvalue as $\overline{W}_j$ to distinguish from the operator $W_j$. One can prove, as in Appendix A of~\cite{buca2012}, that at least one steady-state solution exists for each choice of eigenvalue. 

\begin{figure}[tb]
  \centering
  \includegraphics[width=0.45\textwidth]{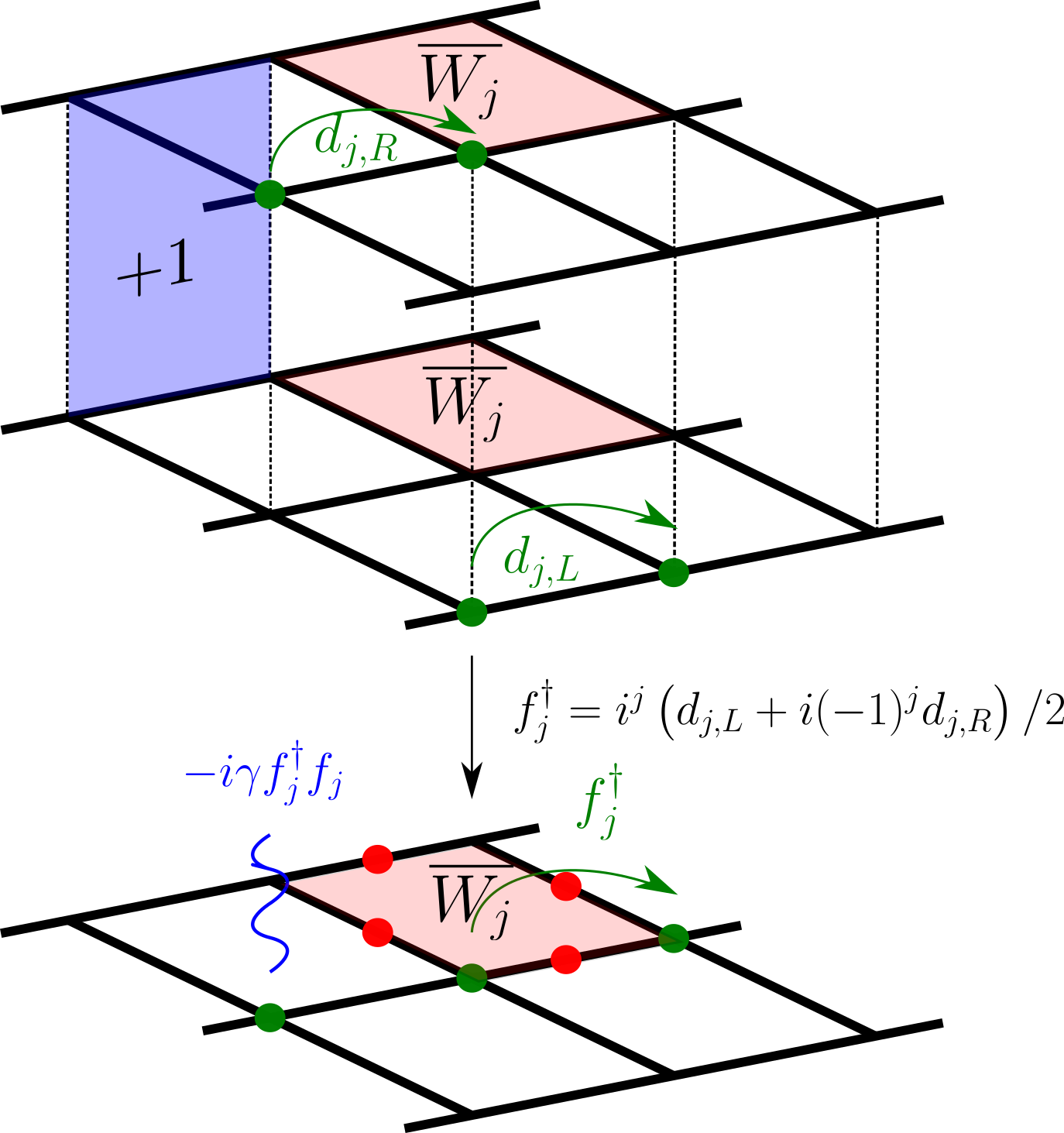}
  \caption{In the steady-state gauge sector, where all interlayer fluxes are set to $+1$ and intralayer fluxes are equal on the two layers, the bilayer Majorana Lindbladian can be directly mapped onto a model of complex fermions on a square lattice coupled to a $\mathbb{Z}_2$ gauge field on the links and with an imaginary chemical potential.}
  \label{fig:bilayerTrans}
\end{figure}

Translating the above statements to our gauge field representation, we fix our gauge sector to be $\hat{w}_{j, \alpha, R} = \hat{w}_{j, \alpha, L} \equiv \hat{w}_{j, \alpha}$ and $\hat{v}_j = (-1)^{j}$. The complex fermion representation chosen in Eq.~\ref{eq:complexRep} makes the Lindbladian in the steady-state gauge sector especially simple, as 
\begin{equation}
  \begin{aligned}
  2 (f^\dagger_{j} f^\pdagger_{j + \hat{x}} + f_{j +\hat{x}}^\dagger f_j^\pdagger) &= i d_{j, R} d_{j + \hat{x}, R} - id_{j, L} d_{j + \hat{x}, L} \,,
    \\
    2 f_j^\dagger f_j^\pdagger &= 1 - (-1)^j i d_{j, R} d_{j, L}\,,
  \end{aligned}
\end{equation}
where an identical relation as in the first line but for $\hat{x} \leftrightarrow \hat{y}$ also holds. As a consequence, the Lindbladian takes the simple form
\begin{equation}
  \begin{aligned}
    i \mathcal{L} &= \sum_j \left( J_x \hat{w}_{j, \hat{x}} f_j^\dagger f_{j + \hat{x}}^\pdagger + J_y \hat{w}_{j, \hat{y}} f_j^\dagger f_{j + \hat{y}}^\pdagger + \text{h.c} \right) 
    \\
    &- 2 i\gamma \sum_j f_j^\dagger f_j^\pdagger,
    \label{eq:hoppingStructure}
  \end{aligned}
\end{equation}
see Fig.~\ref{fig:bilayerTrans}.
The non-Hermiticity of $i \mathcal{L}$ is manifest as simple imaginary chemical potential, and we can immediately identify the steady-state solution as the $f_j^\dagger$ vacuum state. The real part of the dispersion is unaffected by the dissipation, and all excitations come with the same dissipative energy penalty $2 \gamma$.

What are the expectation values of observables in these steady-state solutions? Recall that these solutions have eigenvalue $1$ under the symmetries $U_{j\alpha}$ and $W_{jR} W_{jL}$. Any observable with a non-zero expectation value with respect to this steady-state must have identical eigenvalues. Phrased in terms of operators on our original Hilbert space, the requirement is that observables must commute with the flux operators $W_j$ and the bond operators $V_{j\alpha}'$. This is a strong constraint - the only operators that satisfy this condition are precisely products of the $V_{j\alpha}$ bond operators defined in Eq.~\ref{eq:bondOperators1}. One can check explicitly that these operators satisfy the required constraints, and the claim that these are the only operators with such a property follows from dimension counting, worked out in Appendix~\ref{app:fermionOperators}. Physically, these correspond to all operators that can be expressed in terms of pairs of $d_j$ Majorana fermions connected by strings of $\mathbb{Z}_2$ gauge fields $\hat{w}_{j \alpha}$. 

We now argue that among these operators, only \textit{closed loops} of $V_{j\alpha}$ operators have a non-zero expectation value - recall that these correspond to products of flux operators $W_j$. This is a consequence of the steady-state solution being the vacuum state of the $f_j^\dagger$ operators, which gives an additional set of constraints: $(1 - 2 f_j^\dagger f_j) \Vert \rho \rAngle = \Vert \rho \rAngle$. We can turn this into a gauge-invariant statement by the following rewriting
\begin{equation}
  \begin{aligned}
     \Vert \rho \rAngle&= (-1)^j (1 - 2 f_j^\dagger f_j) \hat{v}_j \Vert \rho \rAngle
    \\
    &=   d_{j, R} d_{j, L} d_{j, R}' d_{j, L}' \Vert \rho \rAngle 
    \\
    &= \Gamma_{j,R}^5 \Gamma_{j,L}^5 \Vert \rho \rAngle \,. 
    \label{eq:vacuumConstraint}
  \end{aligned}
\end{equation}
Hence, any non-zero observable must have eigenvalue $1$ under the symmetry $\Gamma^5_{j, R} \Gamma^5_{j, L}$ (i.e., they commute with $\Gamma^5_j$), and these are precisely closed loops of $V_{j, \alpha}$ operators. Using the fact that the steady-state solution obeys the relation $W_{j, L} \Vert \rho \rAngle = W_{j, R} \Vert \rho \rAngle \equiv \overline{W}_j \Vert \rho \rAngle$, we can deduce that the expectation value of the flux operators in this steady state are given precisely by the intralayer gauge fluxes $\overline{W}_j$. 

\begin{figure}[tbp]
  \centering
  \includegraphics[width=0.45\textwidth]{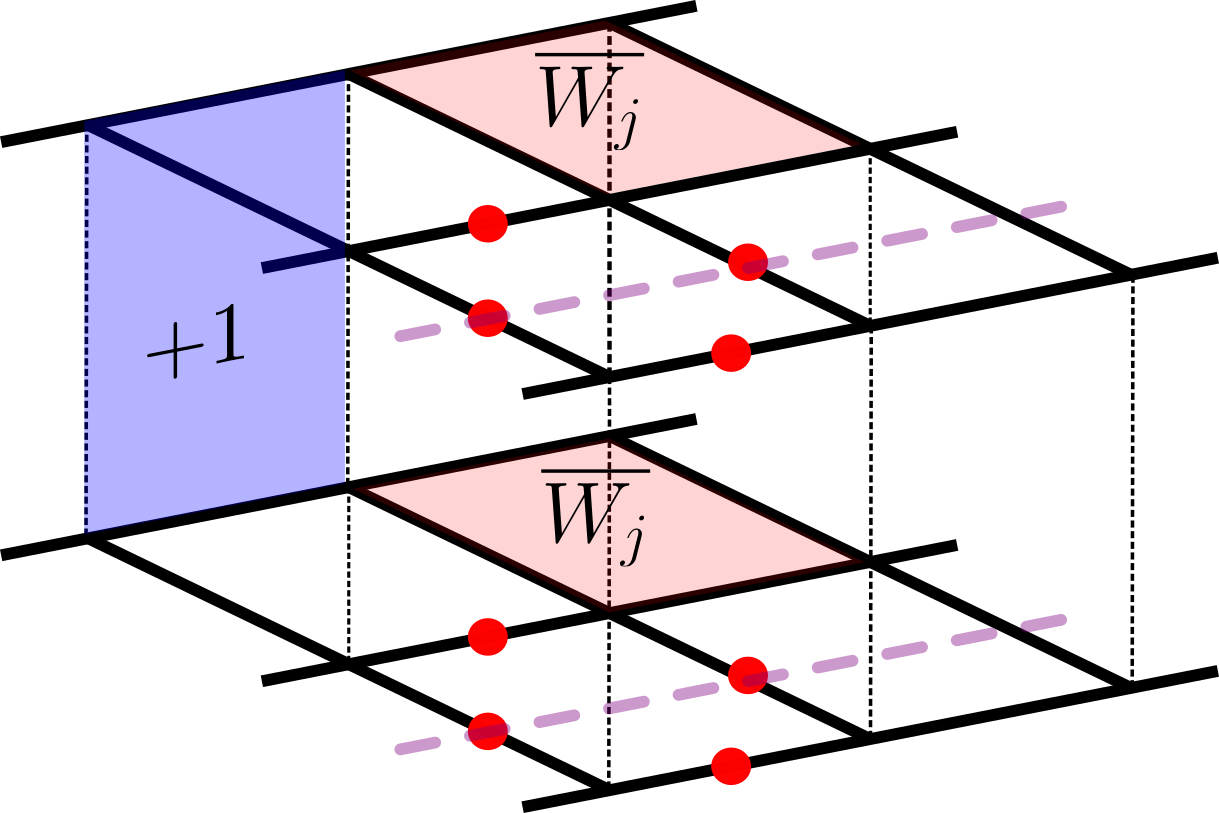}
  \caption{When our model is defined on a torus, one can flip non-contractible loops of intralayer gauge fields in order to obtain a set of four steady-state solutions with equal flux configurations.}
  \label{fig:wilsonLoop}
\end{figure}

When our model is defined on a torus, the steady states of our Lindbladian exhibit a four-fold topological degeneracy arising from the possibility of flipping non-contractible loops of $\hat{w}_{j, \alpha}$ operators, shown in Fig.~\ref{fig:wilsonLoop}. Physically, this implies four distinct steady-state density matrices $\rho_{1-4}$ for each local flux configuration, which are distinguishable based on the expectation values of non-contractible strings of $\Gamma$ matrices. We emphasize that, while this may be thought of as a topological degeneracy - and more generally, $\mathbb{Z}_2$ topological order - within the doubled Hilbert space formalism, it does not constitute true mixed state topological order in the sense of being able to encode logical qubits in the steady-state solutions. What may appear to be a ``quantum'' superposition of different topological sectors $\Vert \rho_1 \rAngle + \Vert \rho_2 \rAngle$ within the doubled Hilbert space formalism translates to a mere classical superposition of density matrices $\rho_1 + \rho_2$ within our original Hilbert space (moreover, the relative phase between the superposition of the two steady-states is not freely tunable - it is fixed by the Hermiticity and positive semi-definite constraint on the physical density matrix).

\subsection{Liouvillian gaps}
Moving beyond steady-state solutions, we can calculate the Liouvillian gap - the energy of the next-lowest state in imaginary energy. It is useful to draw a distinction between different types of Liouvillian gaps. The three types of degrees of freedom in our Lindbladian are complex fermions $f_{j}$, interlayer gauge fields $\hat{v}_j$, and intralayer gauge fields $\hat{w}_{j, \alpha, R}$, $\hat{w}_{j, \alpha, L}$. Excitations with respect to any of these three variables may be considered. Recall from Eq.~\ref{eq:gaugeInvariance} that gauge invariance requires an even number of excitations.

\begin{itemize}
  \item Within a gauge field configuration with a steady-state solution, we compute the \textit{fermion gap}, which is the energy associated with a fermionic excitation. In accordance with the condition of gauge invariance discussed previously, any valid state must include a pair of these excitations. 
  \item We also compute the effects of interlayer gauge excitations, which corresponds to the energy associated with flipping a single $\hat{v}_j$ away from the ``checkerboard'' sector. We call this the \textit{interlayer gauge gap}.
  \item Finally, we analyze intralayer gauge field excitations, which come from flipping a single $\hat{w}_{j, L}$ operator. We choose left gauge fields for concreteness - an identical calculation follows for right gauge fields.
\end{itemize}
We will study each of these excitations in turn. In addition to calculating their Liouvillian gaps, we also identify operators whose equilibration timescales can be upper bounded by these gaps. We make this identification primarily through the symmetry-based analysis outlined previously in Section~\ref{subsec:generalRemarks}. To be precise, each of these excitations will be associated with a particular flux configuration, and the excitations can therefore only have a non-zero overlap with operators whose eigenvalues under the flux superoperators are identical. 
This analysis is robust and can be applied to any excitation; however, for interlayer gauge excitations, we will find that the nature of the fermionic degrees of freedom allows us to say more about the structure of the long-lived excitations.
\subsubsection{Fermion gap}\label{FermionGap}
We first study the Liouvillian gap associated with fermionic excitations within the steady-state gauge sector. As is clear from Eq.~\ref{eq:hoppingStructure}, the fermion gap is always $2 \gamma$, and a pair of these excitations will cost energy $4 \gamma$. As these excitations remain in the same gauge sector, they will still have eigenvalue $1$ under the symmetries $U_{j,\alpha}, W_{j,R} W_{j,L}$. Recalling the relation between $f_j$ and the Majorana fermions in Eq.~\ref{eq:complexRep}, we see that this fermion gap of $4 \gamma$ defines the inverse timescale under which the expectation values of pairs of $d_j$ fermions will asymptote to their steady-state value of zero. The fact that also the Hermitian part of $i\mathcal{L}$, the first line in Eq.~\ref{eq:hoppingStructure}, is quadratic means that the (in general $\hat{w}_{j,\alpha}$ dependent) exact eigenstates of the Lindbladian in the steady-state gauge sector and the time-dependent phases they pick up are characterized by all possible occupation numbers of the $N$ Bloch states of the $f_j$ and their band structure; the associated decay rate is just given by $2\gamma$ times the number of occupied Bloch states.

\subsubsection{Interlayer gauge excitation}
\label{subsec:interlayer}
Creating an interlayer gauge excitation at site $k$ gives us the free fermion Lindbladian
\begin{equation}
  \begin{aligned}
    i \mathcal{L} &= \sum_j \left( J_x \hat{w}_{j, \hat{x}} f_j^\dagger f_{j + \hat{x}}^\pdagger + J_y \hat{w}_{j, \hat{y}} f_j^\dagger f_{j + \hat{y}}^\pdagger + \text{h.c} \right) 
    \\
    &- 2 i\gamma \sum_{j \neq k} f_j^\dagger f_j^\pdagger - 2 i \gamma (1 - f_k^\dagger f_k^\pdagger)\,.
    \label{eq:hoppingStructureInterlayer}
  \end{aligned}
\end{equation}
The structure of the Lindbladian is the same for multiple interlayer gauge excitations - the chemical potential at each site is changed from $f_k^\dagger f_k^\pdagger$ to $(1 - f_k^\dagger f_k^\pdagger)$.
A single one of these flips is not gauge-invariant; one must either flip an additional gauge degree of freedom or add in an odd number of fermions in order to recover a physical excitation. The Liouvillian gap for these excitations must be computed numerically since, as opposed to Eq.~\ref{eq:hoppingStructure}, the Hermitian and anti-Hermitian part of $i\mathcal{L}$ do not commute anymore. However, we can readily see analytically that this gap vanishes in the limit of strong dissipation, $\gamma \rightarrow \infty$. In this limit, we ignore the Hermitian terms in Eq.~\ref{eq:hoppingStructureInterlayer} and we can obtain steady-state solutions by simply placing fermions wherever the imaginary chemical potential is negative (this automatically satisfies the gauge constraint, as we place as many fermions as we flip $\hat{v}_i$'s).

\begin{figure}[tb]
  \centering
  \includegraphics[width=0.5\textwidth]{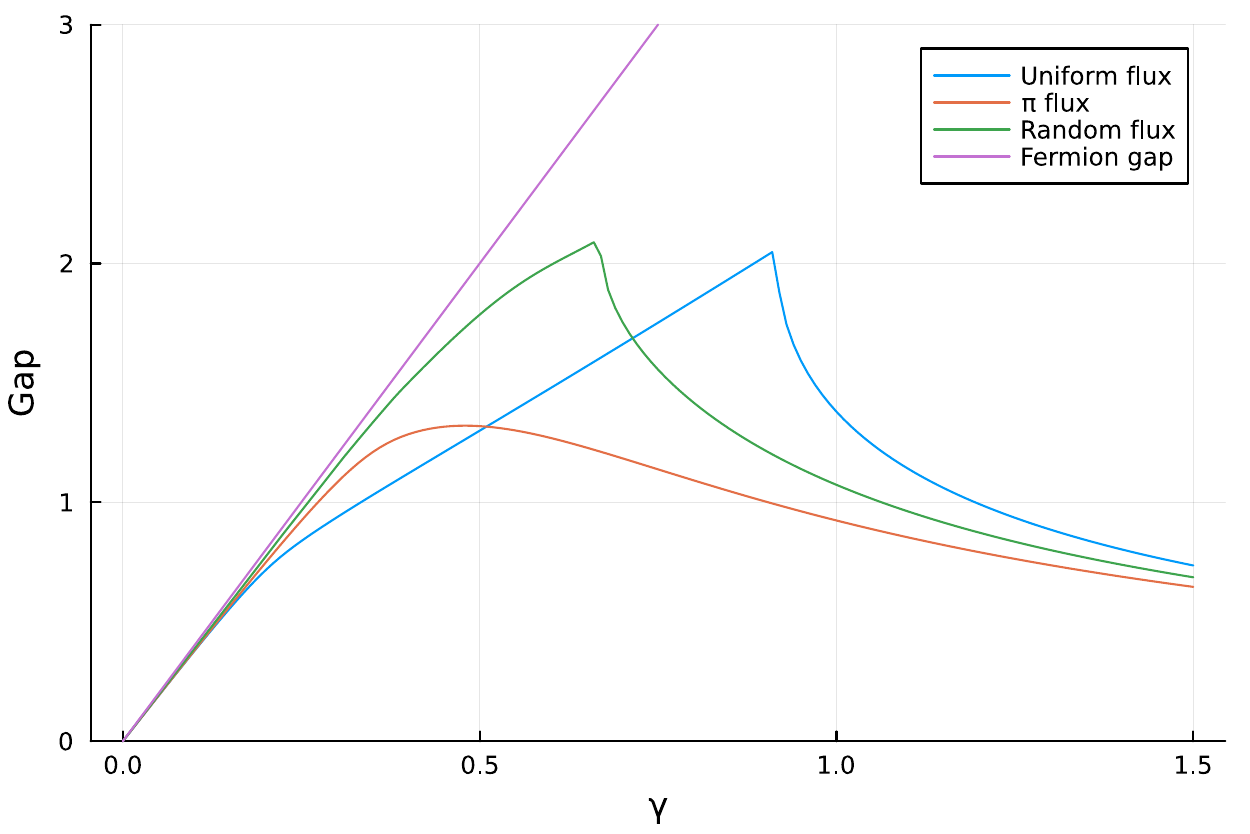}
  \caption{We plot the Liouvillian gap associated with the flipping of a single interlayer gauge degree of freedom $\hat{v}_j$ for various different background flux configurations. All configurations have a quantum Zeno limit as $\gamma \rightarrow \infty$, where the Liouvillian gap vanishes and a new steady-state emerges. For comparison, we also plot the fermion gap of $4 \gamma$ and note that the interlayer gauge gap has an identical slope at small $\gamma$.}
  \label{fig:singleParticleGap}
\end{figure}

For general $\gamma$, the gap of interlayer gauge with fermion excitations (i.e., flipping a single $\hat{v}_k$ and introducing a single fermion to the vacuum) is plotted in Fig.~\ref{fig:singleParticleGap}. For this and all subsequent plots, the parameters used were $J_x = J_y \equiv J = 1$, and $N=1600$. The gap depends on the background $W_j$ flux configuration - we present results for zero flux, $W_j = +1$, $\pi$-flux, $W_j = -1$, and a random flux configuration. Note that there are two distinct contributions to the Liouvillian gap in Eq.~\ref{eq:hoppingStructureInterlayer}. The first is the overall shift of $2i\gamma$, and the second comes from the dissipative strength of the fermion excitation with the smallest imaginary energy. For small $\gamma$, the imaginary energy of this fermion excitation is positive - in other words, adding in the single fermion excitation to the vacuum is energetically unfavorable and causes the eigenstate to decay more rapidly, but one is nevertheless forced to include it by the constraint of gauge invariance. This fermion excitation energy eventually transitions from positive to negative, asymptotically approaching $-2i\gamma$. 

Depending on the background flux configuration, the fermion spectrum may exhibit an anti-$\mathcal{P}\mathcal{T}$-symmetry breaking transition at a critical value of $\gamma$, which causes a sharp kink in the gap. In this situation, the eigenvalues with the smallest imaginary part for small $\gamma$ come in pairs, with the real parts opposite in sign - this symmetry is a consequence of the Lindbladian descending from a completely positive and trace-preserving quantum channel and can be expressed in terms of modular conjugation~\cite{kawabata2023}. The anti-$\mathcal{P}\mathcal{T}$-symmetry breaking transition happens when the two eigenvalues meet on the imaginary axis and split off. We see that in Fig.~\ref{fig:singleParticleGap}, this happens for both the uniform flux as well as the particular random flux configuration plotted, but not for the $\pi$-flux scenario. A survey of generic random flux configurations suggest that this transition is common but not necessarily guaranteed. As this symmetry-breaking transition pertains to dissipative rather than steady-state solutions, the physical consequence of the transition are more subtle, although in principle it may be detected by longest-lived mode in this sector transitioning from having a real (oscillatory) component to being purely dissipative.

What is the physical interpretation of these interlayer gauge excitations? As was the case in the steady-state gauge sector, we can proceed with a symmetry analysis of the operators in this sector. In terms of gauge-invariant fluxes, the flip of a single $\hat{v}_j$ away from its steady-state checkerboard configuration changes the fluxes of the four neighboring $U_{j,\alpha}$ operators to be $-1$. Hence, operators that have a non-zero overlap with this excitation must have identical eigenvalues under these flux operators. Recall that in the steady-state sector, the operators that satisfied the flux constraint consisted of 
pairs of $d_j$ fermion excitations connected by a string of gauge fields $\hat{w}_{j,\alpha}$. An interlayer gauge excitation at site $k$ ``pins'' a $d_k'$ fermion excitation to site $k$, and the allowed operators are gauge-invariant string-like operators that involve a $d_k'$ fermion at site $k$. Therefore, the Liouvillian gap in Fig.~\ref{fig:singleParticleGap} determines the equilibration timescale of operators given by a single $d'$ fermion coupled to a $d$ fermion by a $\mathbb{Z}_2$ Wilson line.

The above argument applies to all operators in this gauge sector, regardless of their energy. In the limit $\gamma \rightarrow \infty$, we can also analytically understand the nature of the lowest-energy (i.e.~the longest lived) operator in this sector. For an interlayer gauge excitation at site $k$, the steady-state solution obeys $f_k^\dagger f_k \Vert \psi \rAngle = 1$ and $f_j^\dagger f_j \Vert \psi \rAngle = 0$ elsewhere. By leveraging this constraint using analogous manipulations as in Eq.~\ref{eq:vacuumConstraint}, we find that this is only satisfied by the operator $\Gamma^5_k$, which can be interpreted as the bound state of a $d$ and $d'$ fermion localized on a single site [cf.~Eq.~\ref{DefinitionOfMajoranas}].  Hence, in the limit $\gamma \rightarrow \infty$, we recover steady-state excitations with definite $\Gamma^5$ eigenvalue. This is a consequence of the quantum Zeno effect; if we interpret the jump operators $L_j = \Gamma^5_j$ as the environment performing measurements of $\Gamma^5$ with frequency specified by $\gamma$, our state can become frozen in a $\Gamma^5$ eigenstate for large $\gamma$. 

\begin{figure}[tpb]
  \centering
  \includegraphics[width=0.2\textwidth]{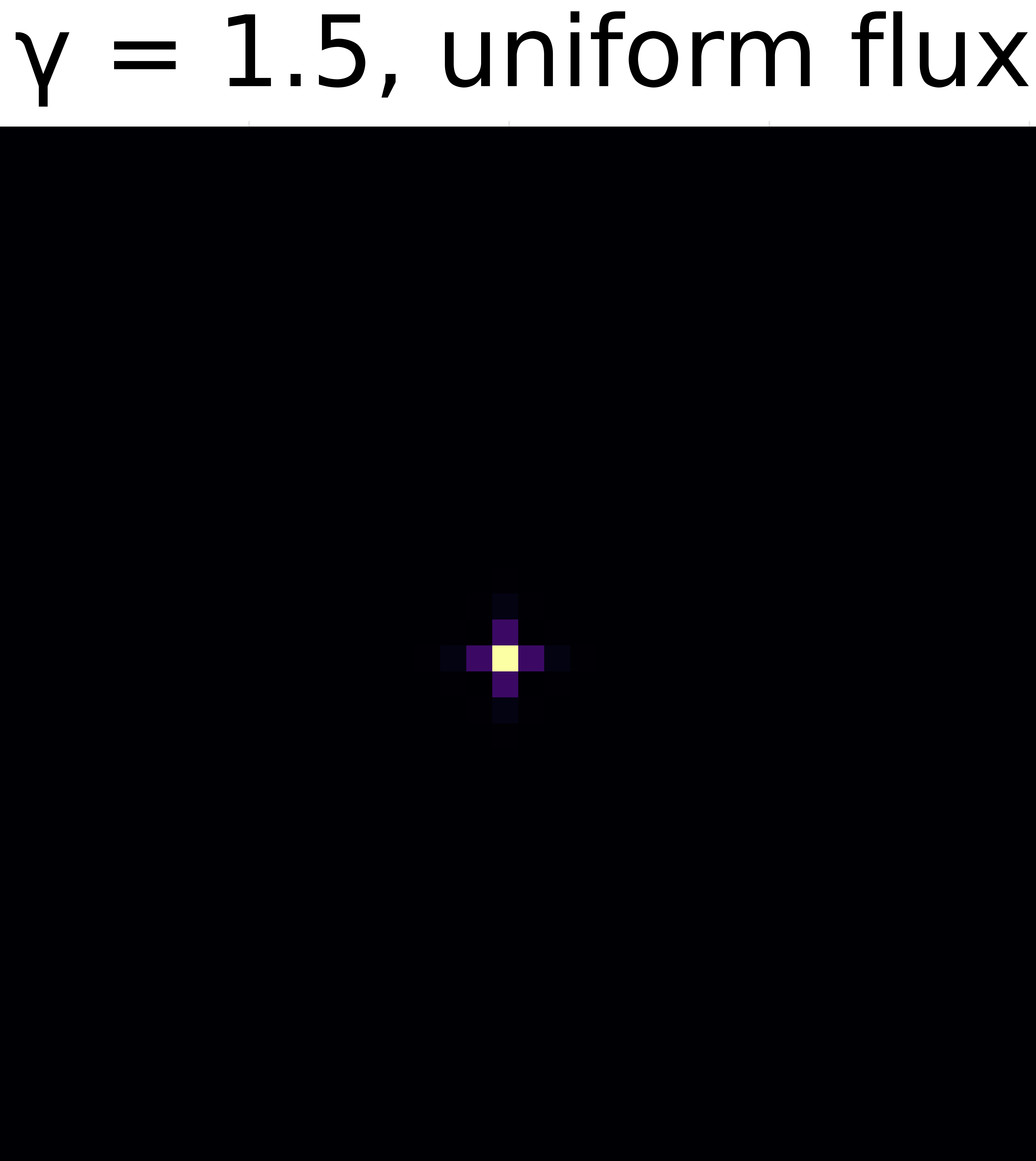}
  \hspace{0.75cm}
  \includegraphics[width=0.2\textwidth]{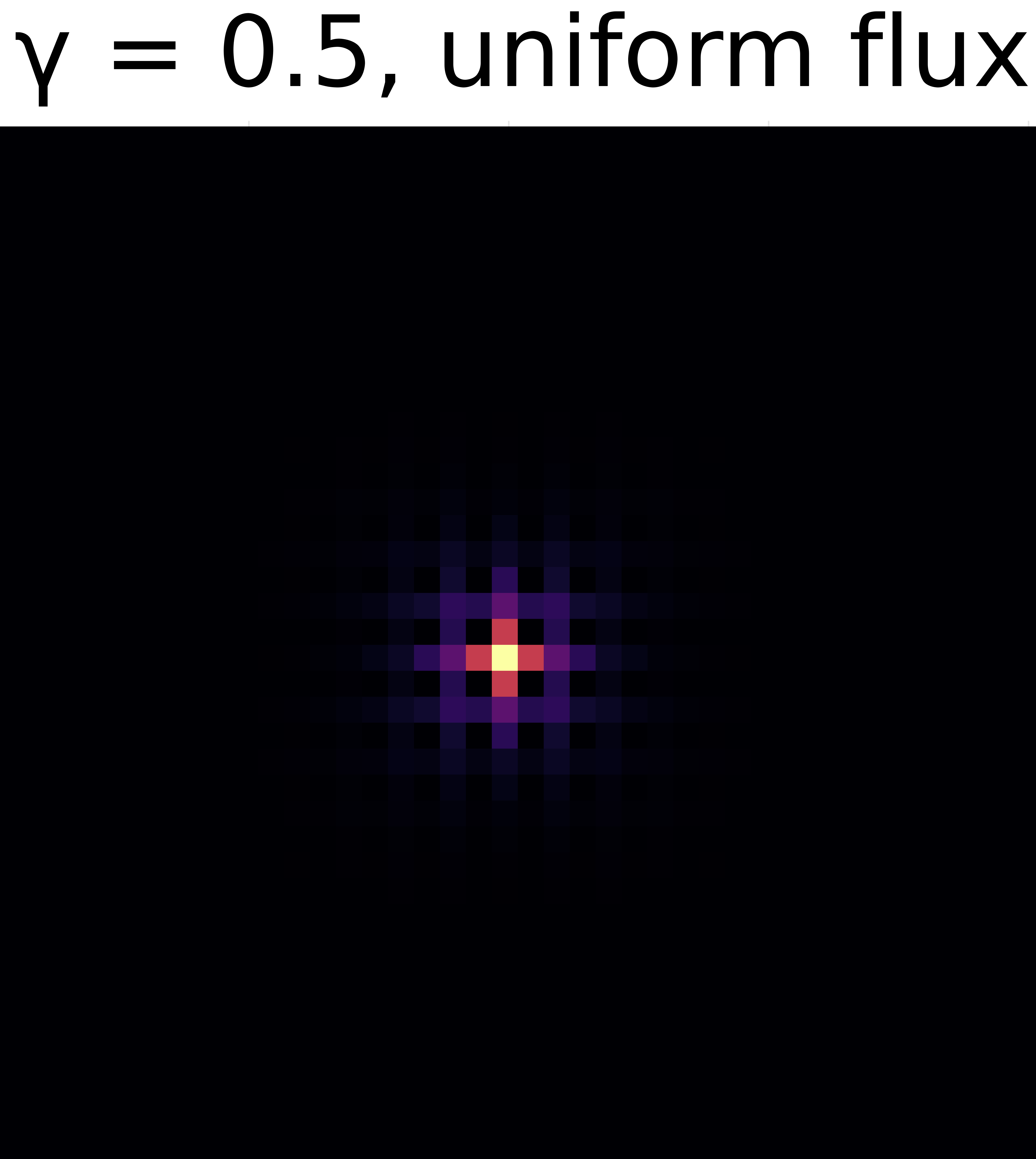}
  \includegraphics[width=0.2\textwidth]{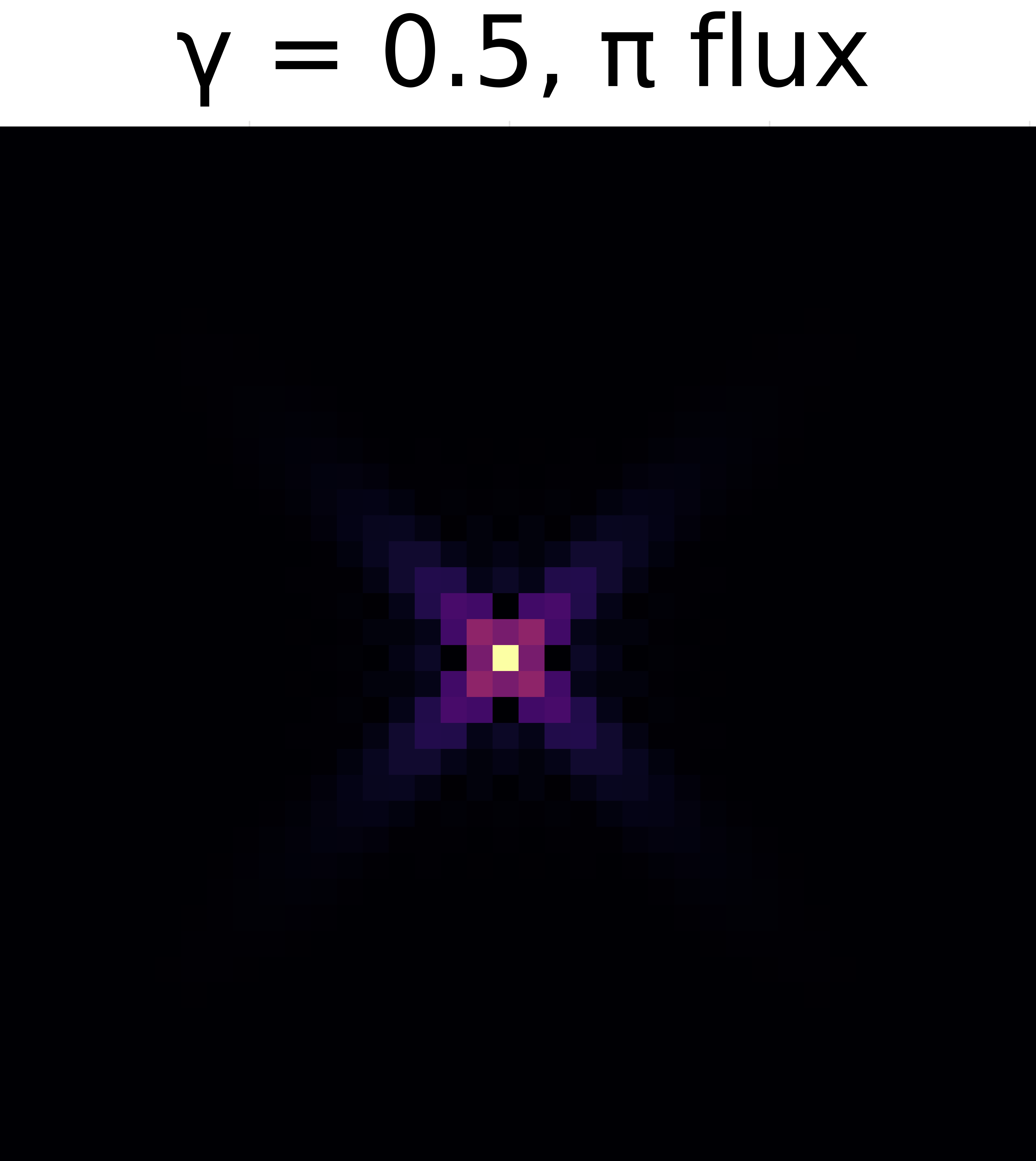}
  \hspace{0.75cm}
  \includegraphics[width=0.2\textwidth]{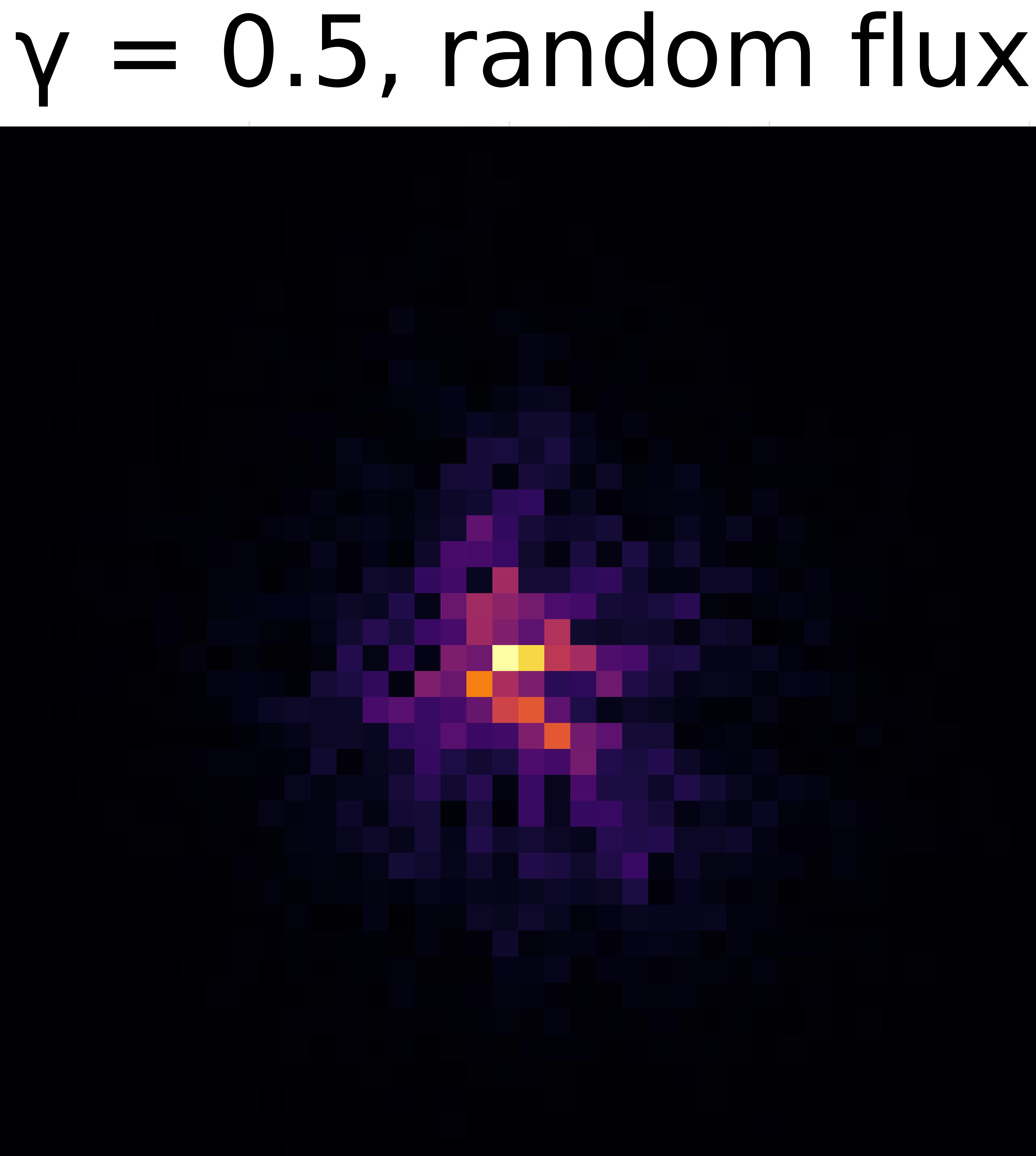}
  \caption{We plot the spatial distribution of the magnitude of the lowest-energy fermion excitation of the single-particle Lindbladian in Eq.~\ref{eq:hoppingStructureInterlayer}, with the single interlayer gauge defect in the center of the lattice. For large $\gamma$ (top left), the eigenstate is highly localized at the defect site, irrespective of the background flux configuration. For smaller $\gamma$, the eigenstate remains well-localized, and the spatial distribution around the defect site is dependent on the background flux.}
  \label{fig:localization}
\end{figure}

The interpretation of the lowest-energy excitation as a $\Gamma^5_k$ operator also holds approximately away from the $\gamma \rightarrow \infty$ limit, which is a consequence of the localization of the corresponding single-particle eigenvector of Eq.~\ref{eq:hoppingStructureInterlayer} around site $k$. As shown in Fig.~\ref{fig:localization}, the fermion with smallest imaginary eigenvalue is highly localized around site $k$ even for small values of $\gamma$; hence, the operator whose equilibration time is determined by Fig.~\ref{fig:singleParticleGap} retains a large overlap with $\Gamma^5_k$. We leave a detailed analysis of the extent of eigenvector localization for future work, although we mention related work~\cite{tzortzakakis2020} of a similar single-particle system but with a fully disordered imaginary chemical potential, rather than our case of a chemical potential that is everywhere positive expect for a single site. For their model, numerical simulations were consistent with a localization transition for arbitrarily weak disorder strength.

The above analysis has been for a single interlayer gauge field excitation. It is natural to consider multiple gauge excitations, which correspond to symmetry sectors with multiple $\hat{v}_j$ gauge fields flipped away from their steady-state configuration. A physically relevant quantity to consider is the Liouvillian gap associated with the $f$ vacuum in the sector with a pair of interlayer gauge field excitations at sites $k$ and $\ell$. This determines the equilibration timescale of an operator given by a pair of $d'$ fermions at sites $k$ and $\ell$. This state is an exact eigenstate of the Lindbladian with imaginary energy $4\gamma$ - note that for sufficiently large $\gamma$, this energy may be reduced further by including pairs of $f$ fermions, with a quantum Zeno effect yielding a steady-state solution at $\gamma \rightarrow \infty$ by adding a pair of fermions at sites $k$ and $\ell$.

\subsubsection{Intralayer gauge excitations}
\label{subsec:intralayer}
The final types of excitation we will study are \textit{intralayer} gauge excitations, when we flip a gauge field on one of the two layers such that $\hat{w}_{k, \alpha, L} = - \hat{w}_{k, \alpha, R}$ for some bond $(k,\alpha)$. Operators associated with these excitations - i.e., operators consistent with this flux configuration - are single-site operators $\Gamma_j^{\mu}$, $\mu = 1, 2, 3, 4$, on the two sites adjacent to the bond $(k, \alpha)$. A more precise identification of these operators, including the flux configurations corresponding to operators $\Gamma^{\mu 5}_j$ and $\Gamma^{\mu \nu}_j$, are given in Appendix~\ref{app:singleSiteOps}. 

\begin{figure}[tpb]
  \centering
  \includegraphics[width=0.5\textwidth]{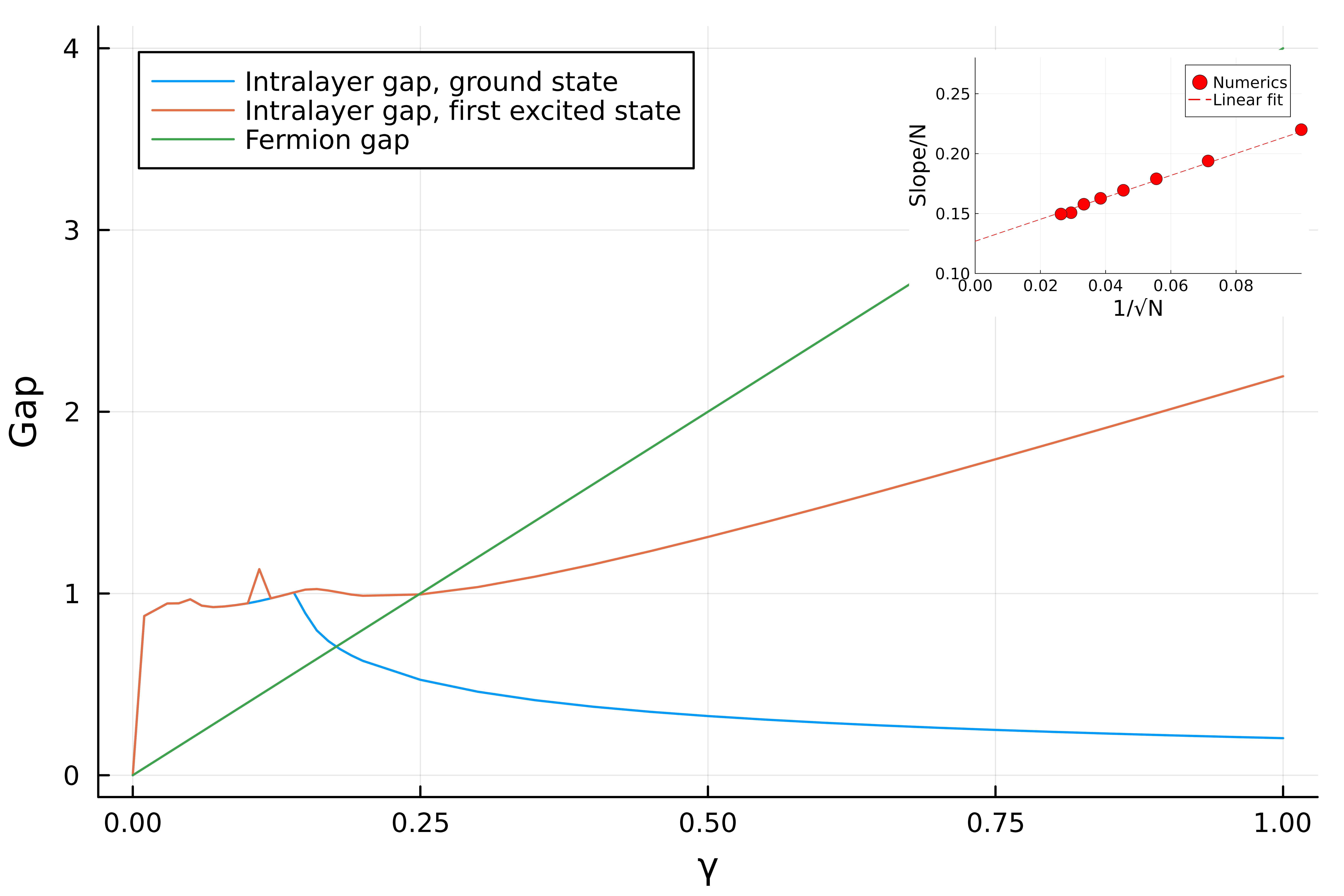}
  \caption{We plot the Liouvillian gap associated with flipping a single intralayer gauge field away from a uniform flux configuration. As it is non-trivial to ensure that the ground state configuration has the proper fermion parity necessary for gauge invariance, we plot both the lowest energy configuration as well as the energy of a state with a single fermionic excitation, with the latter giving a more physical lower bound in the $\gamma \rightarrow \infty$ limit. The Liouvillian gap for small $\gamma$ has a large slope, and a finite size analysis (inset) shows that this slope is extensive in the system size. We plot results for a uniform flux configuration, where the behavior of the gap in the thermodynamic limit appears to take a particularly simple form - an immediate jump up to a gap of magnitude exactly $J$ and a plateau up to $\gamma_c = J / 4$, after which the gap scales linearly in $\gamma$. For generic background flux configurations, we verify that the qualitative nature of the gap remains the same, although the precise coefficients are non-universal.}
  \label{fig:intralayerGaps}
\end{figure}

In this gauge sector, the Lindbladian no longer has a simple expression in terms of complex fermions $f_j^\dagger$, as the intralayer gauge excitation induces pairing terms into the Lindbladian - explicitly, 
\begin{equation}
    2 \left(f_j^\dagger f_{j + \hat{x}}^\dagger + f^\pdagger_{j + \hat{x}} f^\pdagger_j \right) = - i (-1)^j \left(d_{j, L} d_{j + \hat{x}, L} + d_{j, R} d_{j + \hat{x}, R} \right)
\end{equation}
The single-particle Lindbladian is quadratic and can thus still be easily diagonalized; we provide more details of this procedure in Appendix~\ref{app:diag}. However, the determination of whether the resulting ground state is physical - i.e, whether it has the odd fermion parity to not be annihilated by the projection to the physical subspace - is non-trivial due to the non-Hermiticity of the Lindbladian. We leave a full analysis of this problem as an open question and plot both the ground state energy and the energy of the first excited state in Fig.~\ref{fig:intralayerGaps}. The ground state energy gives a lower bound on the physical Liouvillian gap. However, one must be careful at large $\gamma$, since the $\gamma \rightarrow \infty$ limit gives a fictitious quantum Zeno effect. In this limit, the ground state approaches the $f_j$ vacuum state, which is a steady-state solution but unphysical as its fermion parity is even. As a consequence, we also plot the first excited state, which gives a more physical lower bound for large $\gamma$.

We comment on a surprising aspect of this Liouvillian gap, which is a sudden increase when an arbitrarily small $\gamma$ is turned on, with a subsequent plateau at a gap of magnitude $J$. For finite $N$, the gap smoothly evolves as a function of $\gamma$, but the slope at small $\gamma$ is proportional to $N$, as shown in the inset of Fig.~\ref{fig:intralayerGaps}. This indicates that in the thermodynamic limit, an infinitesimally small $\gamma$ causes a discontinuous jump in the Liouvillian gap to $J$. A possible physical explanation of this fact is that, in contrast to the fractionalized operators considered earlier which have a correspondence with coherent excitations of the closed system, the operators $\Gamma_j^{1, 2, 3, 4}$ have no such association, and hence deconstructive interference generated by the unitary dynamics of the closed system also contributes to the decay of the expectation values of these observables. Intuition on this phenomenon can also be gained from the fermion representation - by examining the single-particle eigenstates of the Lindbladian at $\gamma = 0$ expressed in the complex fermion representation, one can see that the act of exchanging a single hopping term with a pairing term causes strong hybridization between the delocalized particle-like and hole-like excitations, which in turn leads to an extensive $\order{\gamma}$ shift in the Liouvillian gap when dissipation is turned on. This phenomenon of the decay rate approaching a non-zero value as $\gamma \rightarrow 0$ in the thermodynamic limit has been found in the Lindladian dynamics of Sachdev-Ye-Kitaev models~\cite{sa2022, kulkarni2022}, 

This observation demonstrates a striking feature of our model in the small-$\gamma$ limit. In this regime, the expectation values of string-like operators such as $V_{j\alpha}$ as well has $\Gamma_j^5$ have an $\order{\gamma^{-1}}$ upper bound on their equilibration timescale, in contrast to local single-site operators such as $\Gamma^{1, 2, 3, 4}_j$ whose timescales are bounded by $\order{J^{-1}}$.

\section{Perturbations away from exact solvability}
\label{sec:perturbations}
As the exact solvability of our Lindbladian requires a precise set of couplings, it is natural to consider perturbations away from this exactly solvable point. Here, we discuss different types of perturbations and their physical effects. Our Lindbladian possesses an extensive number of \textit{strong} symmetries $W_j$ and \textit{weak} symmetries $V_{j,\alpha}$. The combination of the two gives us our exact solvability, and perturbations are conveniently classified in terms of their breaking of these symmetries. 

The simplest perturbations retain both the strong and weak symmetries of our system. These terms are rather artificial - the most local terms consist of either explicitly adding in the flux terms $W_j$ to the Hamiltonian, or adding a two-site jump operator $L_{j,\alpha} = V_{j,\alpha}$. Both these choices preserve the steady-state solutions as well as the structure of the quasiparticle excitations; however, details of the Liouvillian gaps will be modified. 

Perturbations that break the weak symmetries but preserve the strong symmetries of our model include the $J_x'$, $J_y'$, and $J_5$ terms in the full Hamiltonian of Eq.~\ref{eq:squareHamiltonian}. In this case, our quantum jump operators still commute with the fluxes $W_j$, and an initial state in a definite flux sector will remain in that sector for arbitrary time. However, while we can still make statements about the steady-state solutions of the Lindbladian, the full spectrum and consequently the Liouvillian gap is no longer analytically tractable in an exact way. For future work, it would be interesting to study whether coherent quasiparticle excitations still remain in this spectrum at low energies. Recall that in the exactly solvable limit, the existence of distinct types of quasiparticle excitations led to the interpretation of distinct Liouvillian gaps which give equilibration timescales for different observables - the manner in which this picture is modified away from the exactly solvable point is an important open question.

We may also consider perturbations that break the strong symmetries but conserve the weak ones. This is accomplished by a generic choice of quantum jump operator, such as $\Gamma^{1, 2, 3, 4}_j$. In this scenario, we expect our system to asymptote to a unique steady-state, $\rho \propto \id$. The weak symmetries cause the Lindbladian spectrum to decompose into an extensive number of symmetry sectors, with the steady-state solution residing in a particular sector. This means that one still retains the ability to discuss Liouvillian gaps with respect to the steady-state sector versus gaps of different sectors, and a careful analysis of the sectors would allow one to identify the operators that live in these sectors. In passing, we note that particular choices of quantum jump operators such as $\Gamma^{\mu\nu}_j$ with $\mu, \nu \in \{1, 2, 3, 4\}$ will break the local strong symmetries $W_j$ but preserve a global strong symmetry $Q \equiv \prod_j \Gamma^5_j$ (this is not a ``new'' symmetry, as it can be re-expressed as a product of $W_j$ operators). As such, in this case we expect a pair of steady-state solutions $\rho_{\pm} \propto \id \pm Q$.

Finally, a fully generic choice of perturbation that breaks all symmetries will give a single steady-state solution. We again stress an important open question of to what extent quasiparticle excitations of the Lindbladian are robust to these types of perturbations. With regards to the extensive number of steady-state solutions in the exactly solvable limit, one will expect that a small generic perturbation away from this point will cause all but one of these steady-states to persist for a long timescale given by the inverse strength of the perturbation. Developing an analogous theory for the excitations is a promising research direction, as it emphasizes a physical interpretation of the Lindbladian spectrum that is already familiar in the study of closed systems.

\begin{figure}[h!]
  \centering
  \includegraphics[width=\linewidth]{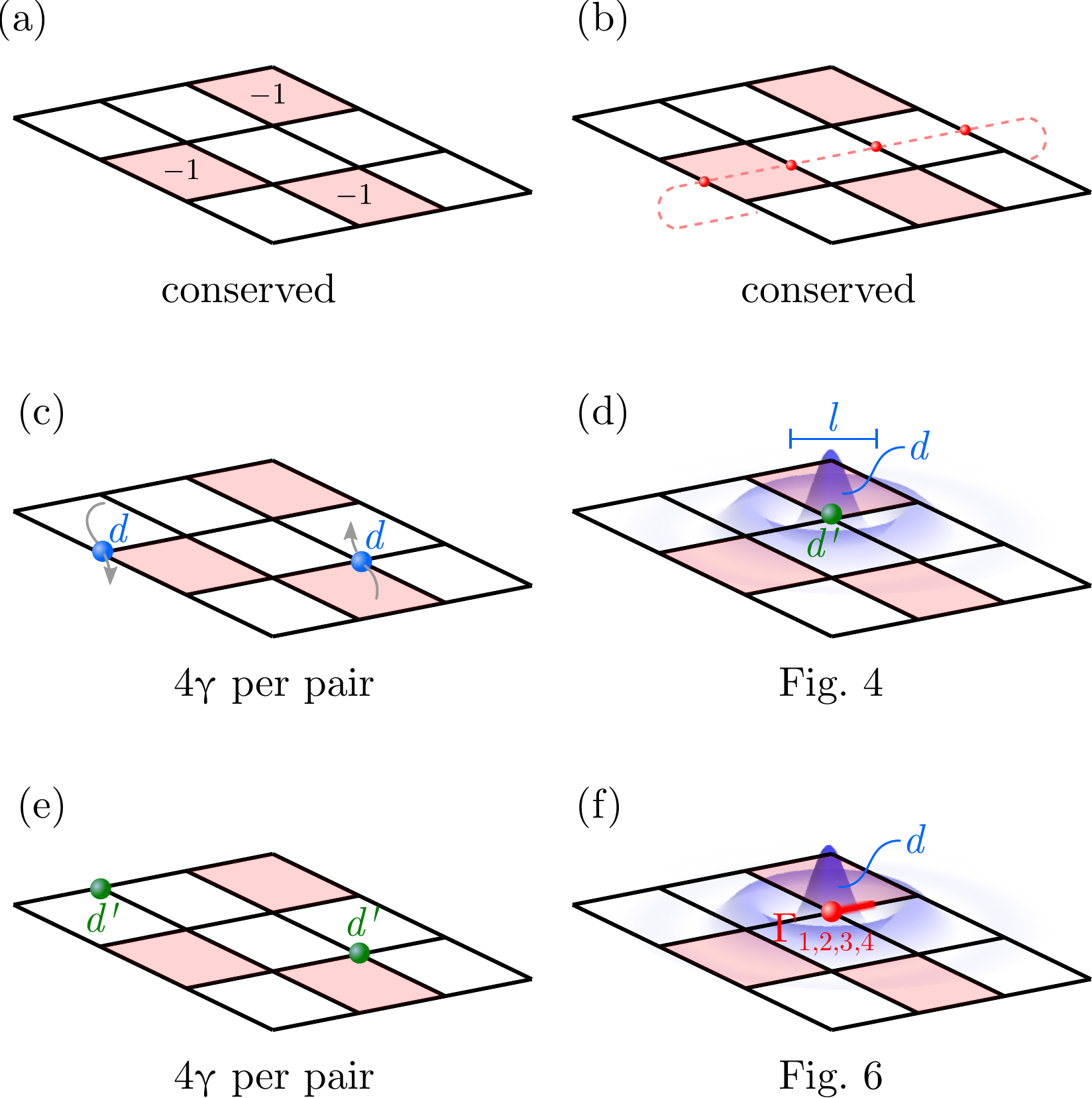}
  \includegraphics[width=\linewidth]{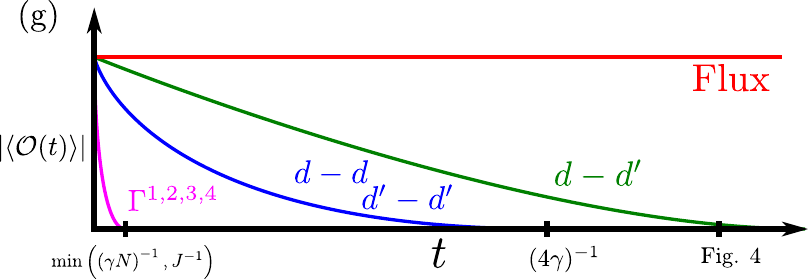}
  \caption{We illustrate the classes of observables considered in our work, as well as their corresponding Liouvillian gaps. a) flux operators $W_j$, which are conserved under Lindbladian time evolution. b) Non-local Wilson loop operators, which are also conserved - we emphasize that \textit{quantum} superpositions of states with different Wilson loop eigenvalues are not steady-state solutions and eventually evolve into classical superpositions. c) A pair of $d$ Majorana fermion excitations, connected by the Wilson line of a $\mathbb{Z}_2$ gauge field (not shown). These have a Liouvillian gap of $4 \gamma$ and the arrows indicate that the eigenstates of $i\mathcal{L}$ are delocalized Bloch-wave-like configurations. d) A $d$ and $d'$ Majorana fermion, connected by the Wilson line of a $\mathbb{Z}_2$ gauge field. The Liouvillian gap is shown in Fig.~\ref{fig:singleParticleGap}, and we find that the wavefunction of the $d$ fermion is highly localized around the $d'$ site. The single-site operator $\Gamma^5_j$ corresponds to the limit $l \rightarrow 0$. e) A pair of $d'$ Majorana fermion excitations, connected by the Wilson line of a $\mathbb{Z}_2$ gauge field. These have a Liouvillian gap of $4 \gamma$. f) Generic single-site operators $\Gamma^{1, 2, 3, 4}$, as well as multi-site operators obtained by including $d$ fermion excitations. Lower bounds on this Liouvillian gap are given in Fig.~\ref{fig:intralayerGaps}. Part g) is a schematic of the resulting decay of the different classes of operators under dissipative time evolution.}
  \label{fig:Excitations}
\end{figure}

\section{Summary and discussion}
\label{sec:summary}
In this work, we analyze the Lindbladian dynamics of a quantum spin-$3/2$ system which admits an exact solution in terms of Majorana fermions coupled to static $\mathbb{Z}_2$ gauge fields. This allows us to characterize the steady-state solutions as well as identify distinct classes of Liouvillian gaps, with different gaps determining the equilibration timescale of different classes of observables, as summarized in Fig.~\ref{fig:Excitations}. Crucially, these timescales fall into different categories with distinct parametric dependencies on $\gamma$. While closed loops of $V_{j\alpha}$ in \equref{eq:bondOperators1}---i.e., the fluxes, Fig.~\ref{fig:Excitations}(a), and on a torus also the non-local Wilson loops, see part (b)---do not decay at all in the exactly solvable limit, pairs of emergent Majorana fermions, Fig.~\ref{fig:Excitations}(c-e), decay with rates that scale linearly with small $\gamma$; depending on whether they exhibit a quantum Zeno effect, these rates decay to zero in the large-$\gamma$ limit. Finally, operators of the last category, like $\Gamma^{1,2,3,4}$, see Fig.~\ref{fig:Excitations}(f), which are not conserved by the Hermitian dynamics, exhibit a decay rate that is singular for small $\gamma$ in the thermodynamic limit; naturally, the entire dynamics is unitary at $\gamma = 0$, however, sending $\gamma \rightarrow 0^+$ \textit{after} taking the thermodynamic limit $N \rightarrow \infty$, the decay rates of these operators is of order of the exchange couplings $J$ of the  Hamiltonian (\ref{eq:squareHamiltonian}). This leads to particularly non-trivial three-step fractionalized thermalization dynamics, see Fig.~\ref{fig:Excitations}(g), in the thermodynamic limit: first, at times of the order of the inverse exchange couplings $1/J$, all operators of the third kind decay, which is parametrically separated from the time-scale $\propto 1/\gamma$ where (gauge invariant pairs of) the Majorana fermions $d$ and $d'$ decay. Then only closed loops of $V_{j\alpha}$ survive, which cannot decay unless perturbations beyond our solvable limit (cf.~Sec.~\ref{sec:perturbations}) are included.

We emphasize that our model, while fine-tuned to ensure exact solvability, demonstrates a novel and potentially generic feature - the presence of distinct quasiparticle excitations in the Lindbladian spectrum, when regarded as a non-Hermitian operator in a doubled Hilbert space, leads to a separation of timescales in the equilibration behavior of different classes of observables. The exact solvability of our model allows us to demonstrate these features explicitly, and can serve as a useful starting point for understanding the robustness of this phenomena in the presence of interactions between quasiparticles. Alternative methods such as the Keldysh formalism~\cite{sieberer2016, yamamoto2021} may prove to be useful in making perturbative treatments of these interactions tractable. 

One promising direction for future research is the construction of additional exactly solvable Lindbladians through this fermionization technique. For closed systems, there exists a rich literature on generalizations of the Kitaev honeycomb model to other exactly solvable models~\cite{yang2007, si2008, wu2009, mandal2009, chua2011}; in these cases, the exact solvability is often geometric in nature (i.e.~arising from a particular choice of lattice connectivity and hopping structure) and is unaffected if a subset of couplings become non-Hermitian. One interesting phenomenon that may arise in a certain parameter regime of these models is \textit{gapless} fermionic excitations, in contrast to our model where fermion excitations have a constant gap $4 \gamma$. This would imply algebraic, rather than exponential, decay of the expectation values of certain classes of operators~\cite{cai2013}. Lindbladians with gapless excitations are not new~\cite{medvedyeva2014, znidaric2015, nakagawa2021, souza2023} - the intriguing new feature of this would be the ability to cleanly separate this spectrum of gapless excitations from gapped gauge excitations, implying distinct equilibration timescales of these operators. 

Generalized Lindbladian constructions may also prove useful at developing a general relation between the exactly solvable open system and the underlying Hermitian dynamics. In our model, the Hermitian dynamics was given by a QSL with two species of Majorana fermions, with the dispersion of one of the fermions tuned to zero. In this limit, a particular choice of quantum jump operators admit quasiparticle excitations of the Lindbladian which display a close relation with the excitation spectrum of the closed system. It is intriguing to ask whether, in a generic system that is rendered exactly solvable through this technique, a similar relation exists between quasiparticle excitations in the doubled Hilbert space and quasiparticle operators of the physical Hilbert space. A more robust understanding of this relation, including potential violations in certain systems, is another promising direction for future research.

\vspace{3em}

\textit{Note added.} Just before posting our work, a related paper appeared on arXiv~\cite{dai2023}, studying exactly solvable BCS-Hubbard Lindbladians. Although the starting point of their analysis involves a distinct microscopic model of complex fermions with pairing terms, a transformation to Majorana fermions yields the same Lindbladian as ours within the $\pi$-flux sector. Due to the different microscopic models, our theory also has a non-trivial gauge invariance requirement, with non-trivial consequences. For instance, the Liouvillian gap in the $\pi$-flux sector in Fig.~\ref{fig:singleParticleGap} is larger as an additional fermion has to be included. We also cite a related work~\cite{gidugu2023} studying a quantum spin Lindbladian very similar to our own, which appeared on arXiv shortly after our posting. Our results are consistent with their analysis.

\begin{acknowledgments}
We thank Pavel Volkov and Hanspeter Büchler for helpful feedback. This work was partially completed at the Center for Computational Quantum Physics in the Flatiron Institute. The Flatiron Institute is a division of the Simons Foundation. L.S acknowledges funding from the U.S. Department of Energy under Grant DE-SC0019030.
\end{acknowledgments}

\appendix
\section{Construction of Gamma matrices}
\label{app:gamma}
In our main text, we outline two possible constructions of Gamma matrices in terms of physical degrees of freedom. There is much freedom in choosing this representation, with different representations making different aspects of the resulting dynamics simpler. An alternate choice is:
\begin{equation}
  \begin{aligned}
    \Gamma^1 &= S^z \otimes S^y \,, \quad \Gamma^2 = S^x \otimes \id \,,  \\
    \Gamma^3 &= S^z \otimes S^x \,, \quad \Gamma^4 = S^y \otimes \id \,, \\
    \Gamma^5 &=S^z \otimes S^z \,. \\
  \end{aligned}
\end{equation}
The unitary dynamics of our model are governed by a Hamiltonian with terms $\Gamma^1_{j} \Gamma^2_{j + \hat{x}}$ and $\Gamma^3_{j} \Gamma^4_{j + \hat{y}}$, which translate into three-spin interactions of the form $S^z_{j, 1} S^y_{j, 2} S^x_{j + \hat{x}, 1}$ and $S^z_{j, 1} S^x_{j, 2} S^y_{j + \hat{y}, 1}$. The jump operator $L_j = S^z_{j, 1} S^z_{j, 2}$ corresponds to a coordinated dephasing term, where the four energy levels of the pair of qubits are subjected to a stochastic noise which leaves fixed the energy difference between the $\ket{\uparrow \uparrow}$ and $\ket{\downarrow \downarrow}$ states, as well as the $\ket{\uparrow \downarrow}$ and $\ket{\downarrow \uparrow}$ states. We note an especially simple feature of this choice, which is the representation of the string-like operators discussed in the main text and whose expectation values decay less rapidly than single-site operators. A string-like operator corresponding to a pair of $d$ Majorana fermion excitations that lies along the $x$-direction is given by $\Gamma^1_j \Gamma^{12}_{j + \hat{x}} \Gamma^{12}_{j + 2\hat{x}} \ldots \Gamma^2_{j + n \hat{x}}$, which in our representation corresponds to a string of $S^y$ operators with an $S^z$ and $S^x$ operator on either end. Similar simplifications arise for strings in the $\hat{y}$ directions, as well as strings corresponding to $d'$ Majorana excitations.
\section{Non-vanishing steady-state expectation values}
\label{app:fermionOperators}
In the main text, we claim that any operator that has eigenvalue $1$ under the superoperators $U_{j,\alpha}$ and equal eigenvalues under $W_{j,R}$ and  $W_{j,L}$ is a product of $V_{j,\alpha}$ bond operators. One can readily verify that these operators satisfy the required constraints, but a more careful argument is required to show that these are the \textit{only} operators with such a property. We do so by counting the dimension of the subspace (within the doubled Hilbert space) spanned by these operators. With a square lattice having $2N$ bonds, there are naively $2^{2N}$ orthogonal combination of bond operators; however, this double counts the true number of operators, as the product of all bond operators is $\id$. So, the subspace is $2^N$ dimensional. The full dimension of our doubled Hilbert space is $2^{4N}$, and we have $3N$ independent constraints - for each site $j$, we have $U_{j,\hat{x}} = 1$, $U_{j,\hat{y}} = 1$, and $W_{j,R} = \bar{W}_j$ (the constraint on $W_{j,L}$ is automatically satisfied under these constraints). Each constraint halves the dimension of the allowed subspace, so we find a $2^N$ dimensional Hilbert space, as desired.

\section{Diagonalization of the free fermion Lindbladian}
\label{app:diag}
In this appendix, we provide more detail on the diagonalization of the free fermion Lindbladian. For a general choice of gauge sector, we work with the Lindbladian written in terms of Majorana fermions, as in Eq.~\ref{eq:freeFermionHamiltonian}. This can be re-expressed in the form
\begin{equation}
    i \mathcal{L} = \bm{d}^T \cdot \bm{A} \cdot \bm{d} - i \gamma N
\end{equation}
where $\bm{d}$ is a $2N$-dimensional vector containing both $d_{j, L}$ and $d_{j, R}$ Majorana fermion operators. We follow the procedure described in~\cite{prosen2008} for obtaining the spectrum of this Lindbladian, which we summarize here. As $\bm{A}$ is an antisymmetric matrix, its spectrum comes in the form $\left\{ \beta_1\,, -\beta_1 \,, \beta_2\,, -\beta_2 \ldots \beta_N\,, -\beta_N\right\}$, where we take $\Im \beta_\alpha \geq 0$. One can construct $N$ creation/annihilation operators $b_\alpha$, $b_\alpha'$ that obey the canonical fermionic anti-commutation relations (with the caveat that $b_\alpha'$ is in general not the Hermitian adjoint of $b_\alpha$). With this, we can write
\begin{equation}
    i \mathcal{L} = - 2 \sum_{\alpha=1}^N \beta_\alpha b_\alpha' b_\alpha - \left(i \gamma N - \sum_{\alpha=1}^N \beta_\alpha\right)
\end{equation}
The term in parenthesis gives the dissipative strength of the state with weakest dissipation within this gauge sector. Note that this Majorana fermion representation obfuscates the constraint of gauge invariance, which is most easily enforced in terms of the complex fermions $f^\dagger_j$. As such, this representation is only useful in gauge sectors where pairing terms would appear if written in the $f_j^\dagger$ basis, in which case a proper analysis of gauge invariance is equally difficult in either representation. 
\section{Identification of single-site operators with flux configurations}\label{app:singleSiteOps}
In the main text, we emphasize that the spectrum of our Lindbladian decomposes into an extensive number of symmetry sectors, each of which is specified by a gauge flux configuration. A Liouvillian gap for each sector can be defined, and one can identify operators - which we remind the reader should be thought of as \textit{states} in this doubled Hilbert space - that are contained in these symmetry sectors, which the Liouvillian gap then defines an equilibration timescale for. Here, we catalog the flux configurations associated with the set of single-site operators.

A particular flux configuration is defined by the interlayer fluxes $U_{j\alpha} = V_{j, \alpha} = V'_{j, \alpha, R} V'_{j, \alpha, L}$ as well as the intralayer fluxes $W_{j, \alpha, R}$, $W_{j, \alpha, L}$. As our Lindbladian spectrum is invariant under the transformation $W_{j, \alpha, R} \leftrightarrow W_{j, \alpha, L}$, we will only identify operators based on their eigenvalues under the combined flux $W_{j, \alpha, R} W_{j, \alpha, L}$. The eigenvalues of an operator under these fluxes is simply determined by whether the operators $V'_{j, \alpha}$ and $W_{j}$ commute or anti-commute with the operators. If we take as our basis of operators to be products of $\Gamma$ matrices, every basis operator will either commute or anti-commute with $V'_{j, \alpha}$ and $W_j$.

The operators $\Gamma_k^5$ commutes with all plaquette operators $W_j$. It also commutes with all the bond operators $V'_{j, \alpha}$ aside from the four bonds adjacent to site $k$. The flux configurations associated with this operator are given precisely by the interlayer gauge excitations studied in Section~\ref{subsec:interlayer}.

The operators $\Gamma_k^\mu$, $\mu = 1, 2, 3, 4$, commute with all the bond operators $V'_{j, \alpha}$ except for a single one adjacent to site $k$ which anticommutes with it. Additionally, it commutes with all but two $W_j$ operators - these two offending plaquette operators share a bond given by the anticommuting $V'_{j, \alpha}$ operator. The flux configuration associated with these operators can be obtained starting from a steady-state gauge sector and flipping an intralayer gauge field on this bond and its spectrum is analyzed in Section~\ref{subsec:intralayer}.

\begin{figure}[tb]
    \centering
    \includegraphics[width=0.5\textwidth]{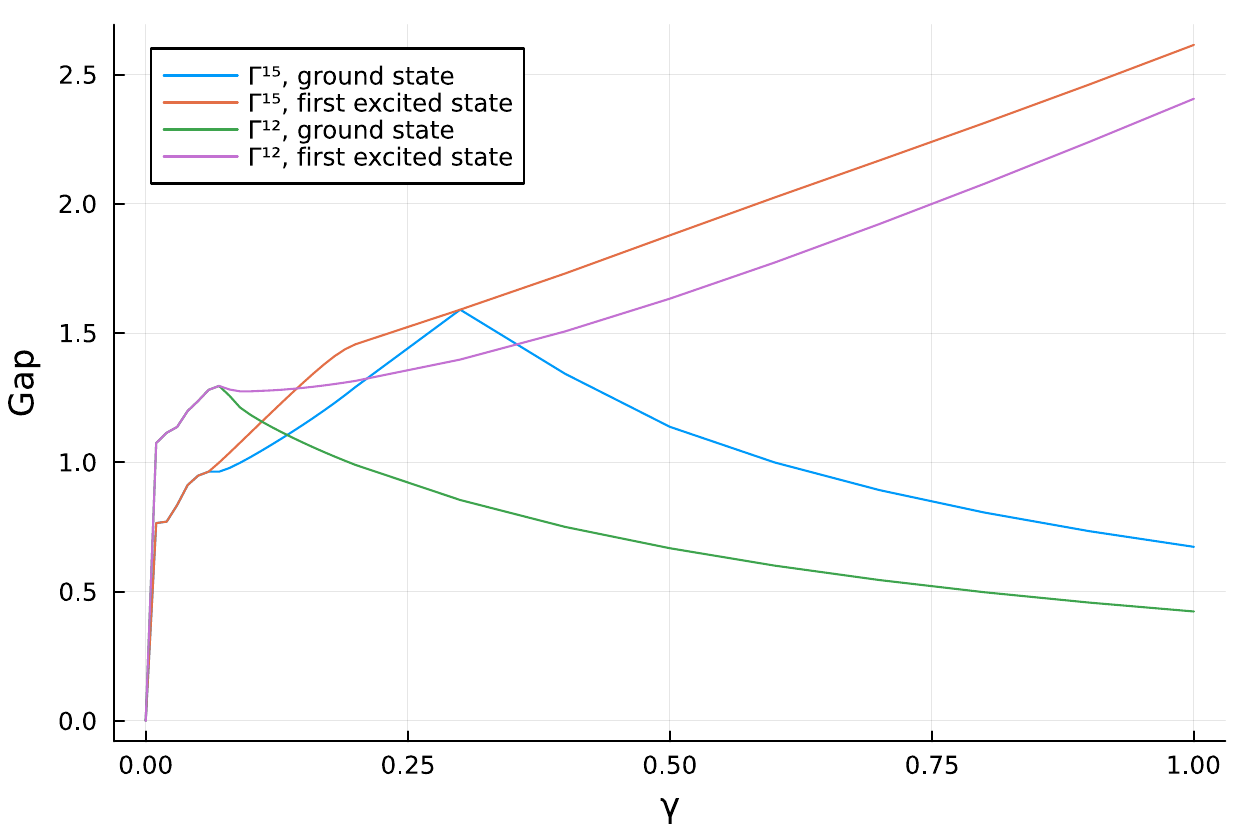} 
    \caption{We plot Liouvillian gaps for the gauge sectors associated with $\Gamma^{15}$ and $\Gamma^{12}$ operators, and demonstrate a sharp jump in the gap when the dissipation is turned on.}
    \label{fig:intralayerAll}
\end{figure}

The operators $\Gamma_k^{\mu 5}$ have the same commutation relations with the plaquette operators as $\Gamma_k^\mu$, but differ with respect to the $V'_{j, \alpha}$ operators; it now anticommutes with the three $V'_{j, \alpha}$ bond operators connected to site $k$ that \textit{aren't} the bond shared by the flux operators. This flux configuration can be obtained from the intraylayer gauge excitation studied in Section~\ref{subsec:intralayer} and flipping an additional interlayer gauge field $\hat{v}_k$. 

Finally, we identify the operators $\Gamma_k^{\mu \nu}$, with $\mu, \nu = 1, 2, 3, 4$ and $\mu \neq \nu$. For a given site $k$, there are $\binom{4}{2} = 6$ different operators of this type. These operators will
anticommute with two of the four $V_{j, \alpha}'$ bond operators, and either two fluxes $W_j$ that only share a corner at site $k$ or all four fluxes connected to site $k$. These flux sectors are obtained by flipping \textit{two} intralayer gauge fields connected to a site $k$ - as expected, there are $\binom{4}{2} = 6$ ways of doing this. 

The Liouvillian gap of excitations corresponding to the $\Gamma_k^\mu$ operators are shown in Fig.~\ref{fig:intralayerGaps}. We plot the Liouvillian gap of $\Gamma^{\mu 5}_k$ and $\Gamma^{\mu \nu}_k$ operators in Fig.~\ref{fig:intralayerAll} and verify that similar behavior occurs. This implies that our observation of the rapid equilibration of $\Gamma^\mu_k$ operators holds \textit{generically} for single-site operators, with the exception of $\Gamma^5_k$ due to its interpretation as the bound state of two Majorana fermion excitations, or alternatively due to the fact that $\Gamma^5_k$ are precisely the quantum jump operators describing the coupling to the environment.

\bibliography{bilayer.bib}
\end{document}